\documentclass[english,british,jcp,reprint,a4paper,prb,superscriptaddress,nobibnotes]{revtex4-1}
\usepackage[T1]{fontenc}
\usepackage[latin9]{inputenc}
\usepackage{amssymb}
\usepackage{graphicx}

\makeatletter

\providecommand{\tabularnewline}{\\}

 \@ifundefined{textcolor}{}
 {%
   \definecolor{BLACK}{gray}{0}
   \definecolor{WHITE}{gray}{1}
   \definecolor{RED}{rgb}{1,0,0}
   \definecolor{GREEN}{rgb}{0,1,0}
   \definecolor{BLUE}{rgb}{0,0,1}
   \definecolor{CYAN}{cmyk}{1,0,0,0}
   \definecolor{MAGENTA}{cmyk}{0,1,0,0}
   \definecolor{YELLOW}{cmyk}{0,0,1,0}
 }

\usepackage[caption=false]{subfig} 

\newcommand*{\citen}[1]{%
  \begingroup
    \romannumeral-`\x 
    \setcitestyle{numbers}%
    \cite{#1}%
  \endgroup   
}

\@ifundefined{showcaptionsetup}{}{%
 \PassOptionsToPackage{caption=false}{subfig}}
\usepackage{subfig}
\makeatother

\usepackage{babel}
\begin{document}

\title{Investigating the significance of zero-point motion in small molecular
clusters of sulphuric acid and water}

\author{Jake L. Stinson}

\email{Email: j.stinson@ucl.ac.uk}

\affiliation{Department of Physics and Astronomy and London Centre for Nanotechnology,
University College London, Gower Street, London, WC1E 6BT, United
Kingdom}

\author{Shawn M. Kathmann}

\affiliation{Physical Sciences Division, Pacific Northwest National Laboratory,
Richland, Washington 99352, United States}

\author{Ian J. Ford}

\affiliation{Department of Physics and Astronomy and London Centre for Nanotechnology,
University College London, Gower Street, London, WC1E 6BT, United
Kingdom}
\begin{abstract}
The nucleation of particles from trace gases in the atmosphere is
an important source of cloud condensation nuclei (CCN), and these
are vital for the formation of clouds in view of the high supersaturations
required for homogeneous water droplet nucleation. The methods of
quantum chemistry have increasingly been employed to model nucleation
due to their high accuracy and efficiency in calculating configurational
energies; and nucleation rates can be obtained from the associated
free energies of particle formation. However, even in such advanced
approaches, it is typically assumed that the nuclei have a classical
nature, which is questionable for some systems. The importance of
zero-point motion (also known as quantum nuclear dynamics) in modelling
small clusters of sulphuric acid and water is tested here using the
path integral molecular dynamics (PIMD) method at the density functional
theory (DFT) level of theory. The general effect of zero-point motion
is to distort the mean structure slightly, and to promote the extent
of proton transfer with respect to classical behaviour. In a particular
configuration of one sulphuric acid molecule with three waters, the
range of positions explored by a proton between a sulphuric acid and
a water molecule at 300 K (a broad range in contrast to the confinement
suggested by geometry optimisation at 0 K) is clearly affected by
the inclusion of zero point motion, and similar effects are observed
for other configurations. 
\end{abstract}
\maketitle

\section{Introduction\label{sec:1Introduction}}

The role of sulphuric acid in the formation of cloud condensation
nuclei (CCN) is believed to be significant\citep{RoyalSocietyGeoEngineeringReport2009,Zhang2012},
on account of its low vapour pressure, relatively high atmospheric
concentration and its affinity to water. However, simple attempts
to understand the binary nucleation of sulphuric acid and water in
detail have proved problematic. It is clear that classical nucleation
theory (CNT) is insufficient for describing this process, since the
critical cluster size suggested from experimental data appears to
be small, and consequently several extensions and alternatives have
been studied \citep{Ford2004,Vehkamaki,Napari2010}.

One approach, the use of atomistic models that explicitly treat individual
molecules or atoms within numerical simulations, has proliferated
as a consequence of increasing computational power; especially based
on quantum chemistry methods which treat the electronic interactions
explicitly. Popular quantum chemistry methods include electronic density
functional theory (DFT) \citep{Re1999,Bandy1998,Ianni2000,Arstila1998,Larson2000,Ding2003,AlNatsheh2004,Arrouvel2005,Ding2004,Nadykto2008,Kurten2009}
and M$\hbox{\o}$ller-Plesset perturbation theory (MPn where n refers
to the order of the perturbation) \citep{Larson2000,Ding2004,Kurten2009,Temelso2012}.
The usual strategy is to identify the lowest energy molecular configuration
and then to use the rigid-rotor-harmonic-approximation (RRHO) to compute
free energies, and thereby investigate cluster stability and nucleation
through specific growth and decay routes.

The Born-Oppenheimer approximation \citep{RichardMMartin2004} is
employed by both DFT and MPn. It involves the separation of the wavefunctions
of electrons and nuclei followed by a classical treatment of the dynamics
of the nuclei. The DFT approach has been used to describe sulphuric
acid and water clusters \citep{Choe2007,Anderson2008,Hammerich2008}.
In simulations based on such approaches, the sulphuric acid and water
system has been observed to exhibit proton transfers. Such events
are of particular importance in this system and a challenge to the
modelling. A question that arises is whether we can account for such
processes correctly while representing the nuclei as classical particles.
Might a quantum treatment of the proton dynamics be more accurate?
Perhaps the additional uncertainty in proton position can alter the
delicate balance between neutral and ionised structures? In this paper
we employ Path Integral Molecular Dynamics (PIMD) to study the quantum
nuclear degrees of freedom (also known as zero-point motion) of sulphuric
acid/water molecular clusters to address this question. A particular
issue for consideration is the level of hydration of a single sulphuric
acid molecule that is required for proton transfer to occur, a matter
that can be addressed either by zero temperature calculations or dynamics
performed at finite temperature. It has been suggested that the threshold
is around three or more water molecules\citep{Arrouvel2005}. Transfer
of the second proton was studied by Ding and Laasonen \citep{Ding2004}
who concluded that it is likely for a level of hydration of around
eight or nine water molecules.

PIMD emulates the quantum behaviour of a particle by using a classical
quasiparticle or bead description, a detailed derivation of which
is given by \citet{MarkE.Tuckerman2010}. The PIMD method has been
shown to have a significant effect on the properties in some hydrogen
bonded systems \citep{Li2011b,Walker2010}. PIMD has been employed
previously together with a parametrised version of the PM6 model \citep{Stewart2007}
(a semi-empirical model of electronic structure) to study sulphuric
acid and water clusters \citep{Kakizaki2009,Sugawara2011}. \citet{Kakizaki2009}
concluded that the PIMD technique (using the normal mode transformation
\citep{MarkE.Tuckerman2010}) increased thermal fluctuations and produced
more liquid-like behaviour in systems at a temperature of ${\rm 250\: K}$
\citep{Kakizaki2009}. \citet{Sugawara2011} studied the degree of
hydration required for the first and second ionisation events for
the sulphuric acid molecule, and concluded that the first ionisation
takes place when four water molecules are present in the cluster in
agreement with earlier work \citep{Arrouvel2005}. The second ionisation
event occurred in the presence of ${\rm 10-12}$ water molecules in
contrast with the study by Ding and Laasonen \citep{Ding2004} though
the latter was based on geometry optimisation techniques rather than
on molecular dynamics. As the purpose of this paper was to gauge the
importance of zero-point motion in the sulphuric acid and water system
as accurately as possible, it was decided to use DFT rather than the
semi-empirical PM6 model developed by \citet{Kakizaki2009} .

We study the importance of zero-point motion in a small cluster of
sulphuric acid and water using PIMD \citep{Feynman2010,MarkE.Tuckerman2010}
as implemented in the CASTEP code\citep{Clark2005}. Section \ref{sec:2Methods}
describes the theory used, section \ref{sec:3Results} details our
results, and section \ref{sec:4Conclusion} concludes our study where
we comment on the significance of zero-point motion in the sulphuric
acid and water system.

\begin{figure}
\begin{centering}
\includegraphics[width=1\columnwidth]{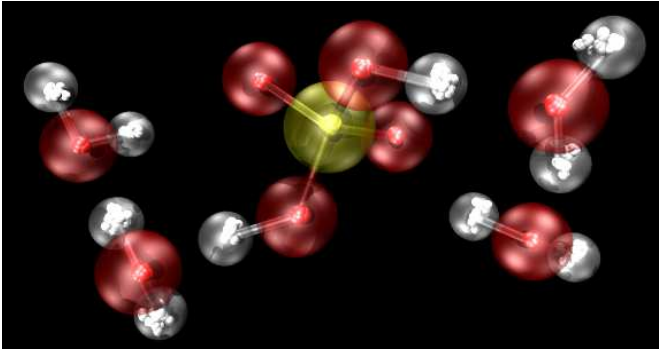}
\par\end{centering}

\caption{\label{fig:Bead picture}A 16 bead representation of a system containing
one sulphuric acid and four water molecules: the distribution of bead
positions conveys the quantum uncertainty. }
\end{figure}

\section{Methods\label{sec:2Methods}}

According to the PIMD technique each particle (nucleus) is represented
by a set of quasiparticles (known as beads) connected by harmonic
springs. The following Hamiltonian describing the bead dynamics can
be derived using the Trotter approximation \citep{MarkE.Tuckerman2010}:
\[
\mathcal{H}(x_{k},p_{k})=\sum_{k=1}^{P}\left[\frac{p_{k}^{2}}{2m'}+\frac{1}{2}m\omega_{P}^{2}(x_{k+1}-x_{k})^{2}+\frac{1}{P}U(x_{k})\right]
\]

\noindent under the condition $x_{P+1}=x_{1}$, where $P$ is the
number of beads and $x_{k}$ and $p_{k}$ are respectively the position
and momentum of bead $k$. $\omega_{P}$ is the harmonic frequency
of the inter-bead springs and is given by $\sqrt{P}/\beta\hbar$ where
$\beta=(k_{B}T)^{-1}$ and $T$ and $k_{B}$ are the system temperature
and the Boltzmann constant respectively. While $m$ denotes the mass
of the particle, the mass of the beads is represented by $m'$, and
$U(x_{k})$ is the classical potential in which the particle moves.
The quantum nuclear behaviour is reflected in both the position and
momentum of the beads under the influence of this Hamiltonian, which
is controlled by the stiffness of the inter-bead springs. Since the
latter is proportional to the mass of the particle, the hydrogen nucleus
is expected to be the most susceptible to zero-point effects.

\begin{table}
\begin{centering}
\begin{tabular}{|c|r@{\extracolsep{0pt}.}l|r@{\extracolsep{0pt}.}l|}
\hline 
 & \multicolumn{2}{c|}{Time step {[}${\rm fs}${]}} & \multicolumn{2}{c|}{Simulation time {[}${\rm ps}${]}}\tabularnewline
\hline 
SATH ${\rm 1}$ bead & \multicolumn{2}{c|}{${\rm 1.00}$} & \multicolumn{2}{c|}{${\rm 11.000}$}\tabularnewline
\hline 
SATH ${\rm 4}$ bead & \multicolumn{2}{c|}{${\rm 0.50}$} & \multicolumn{2}{c|}{$\,\,\,{\rm 1.500}$}\tabularnewline
\hline 
SATH ${\rm 8}$ bead & \multicolumn{2}{c|}{${\rm 0.25}$} & \multicolumn{2}{c|}{$\,\,\,{\rm 0.875}$}\tabularnewline
\hline 
SATH ${\rm 16}$ bead & \multicolumn{2}{c|}{${\rm 0.50}$} & \multicolumn{2}{c|}{${\rm 10.673}$}\tabularnewline
\hline 
SATH ${\rm 32}$ bead & \multicolumn{2}{c|}{${\rm 0.50}$} & \multicolumn{2}{c|}{$\,\,\,{\rm 0.512}$}\tabularnewline
\hline 
SAQH ${\rm 1}$ bead & \multicolumn{2}{c|}{${\rm 1.00}$} & \multicolumn{2}{c|}{$\,\,\,{\rm 1.000}$}\tabularnewline
\hline 
SAQH ${\rm 4}$ bead & \multicolumn{2}{c|}{${\rm 0.50}$} & \multicolumn{2}{c|}{$\,\,\,{\rm 1.500}$}\tabularnewline
\hline 
SAQH ${\rm 8}$ bead & \multicolumn{2}{c|}{${\rm 0.50}$} & \multicolumn{2}{c|}{$\,\,\,{\rm 1.500}$}\tabularnewline
\hline 
SAQH ${\rm 16}$ bead & \multicolumn{2}{c|}{${\rm 0.50}$} & \multicolumn{2}{c|}{$\,\,\,{\rm 1.500}$}\tabularnewline
\hline 
config H ${\rm 1}$ bead & \multicolumn{2}{c|}{${\rm 1.00}$} & \multicolumn{2}{c|}{${\rm 10.900}$}\tabularnewline
\hline 
config H ${\rm 16}$ bead & \multicolumn{2}{c|}{${\rm 1.00}$} & \multicolumn{2}{c|}{${\rm 10.647}$}\tabularnewline
\hline 
\end{tabular}
\par\end{centering}

\caption{\label{tab:timeStepData}Compilation of the simulation length and
time step for the MD runs performed. SATH refers to sulphuric acid
trihydrate and SAQH refers to sulphuric acid tetrahydrate, structures
that assume typical configurations shown in Figure \ref{fig:SATH and SAQH configuration(A)}
and \ref{fig:SATH and SAQH configuration(b)} respectively. Config
H refers to the trihydrate configuration shown in Figure \ref{fig:Config H label}.
Note that the longest simulations were performed for SATH and config
H.}
\end{table}

Figure \ref{fig:Bead picture} is a snapshot from a 16 bead simulation
representing the behaviour of a cluster of one sulphuric acid and
four water molecules. The spatial separation of the beads clearly
illustrates the greater positional uncertainty of the hydrogen nuclei
compared to that of the oxygen and the sulphur nuclei.

\begin{figure}
\noindent \begin{centering}
\subfloat[\label{fig:SATH and SAQH configuration(A)}]{\noindent \begin{centering}
\includegraphics[width=0.48\columnwidth,height=5.6cm]{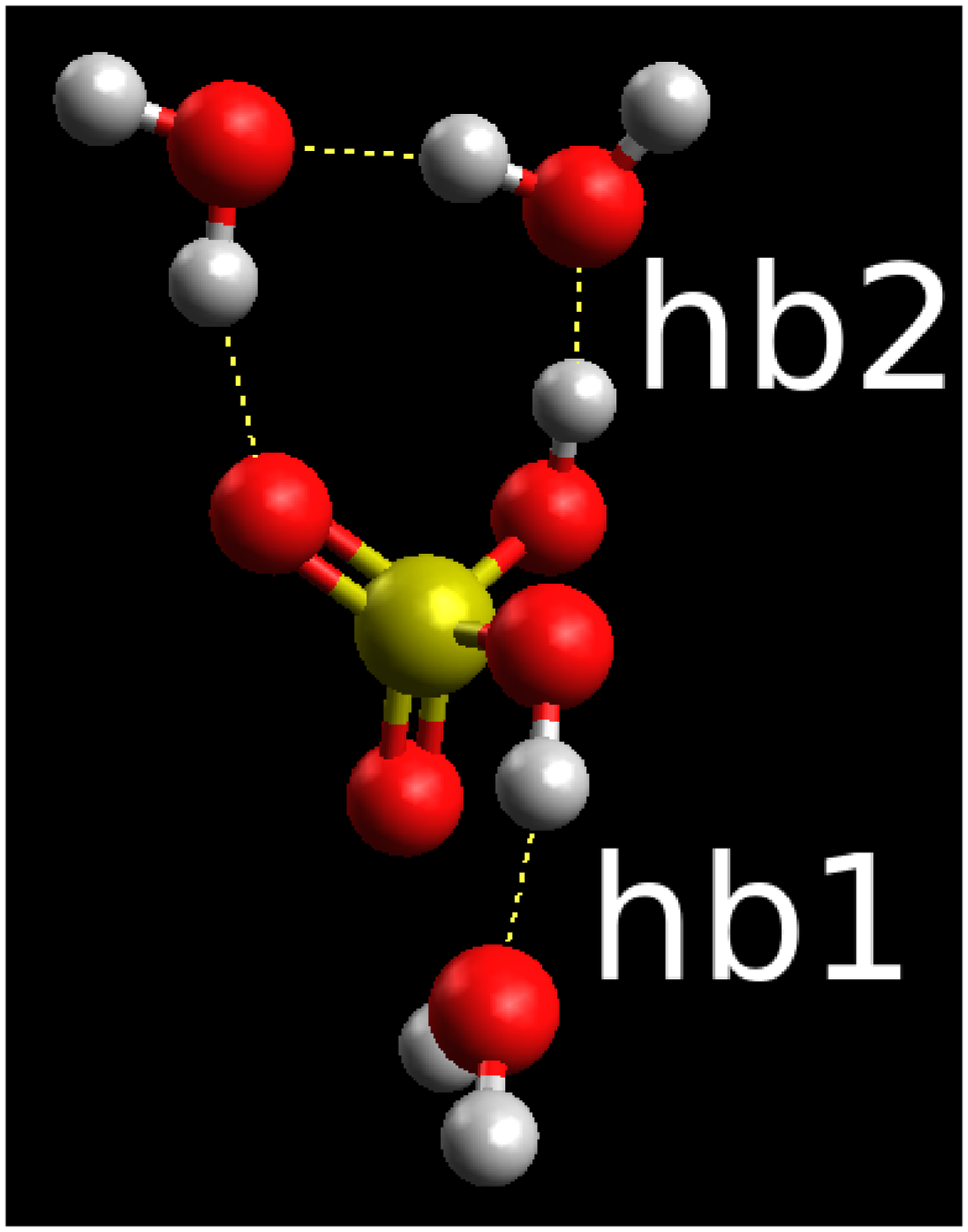}
\par\end{centering}

}\subfloat[\label{fig:SATH and SAQH configuration(b)}]{\noindent \begin{centering}
\includegraphics[width=0.48\columnwidth,height=5.6cm]{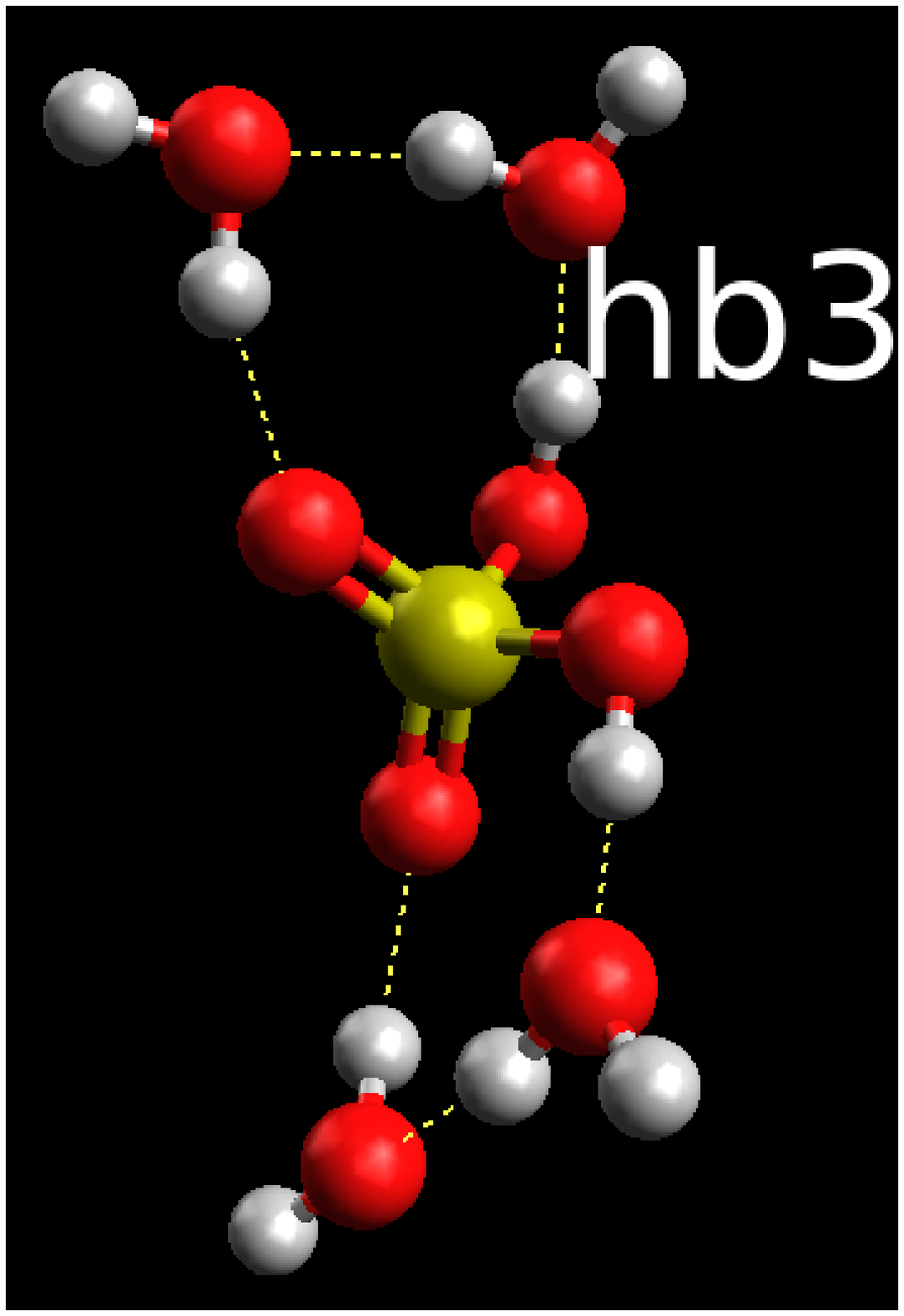}
\par\end{centering}

}
\par\end{centering}

\begin{centering}
\subfloat[\label{fig:binding energy}]{\noindent \begin{centering}
\includegraphics[width=1\columnwidth]{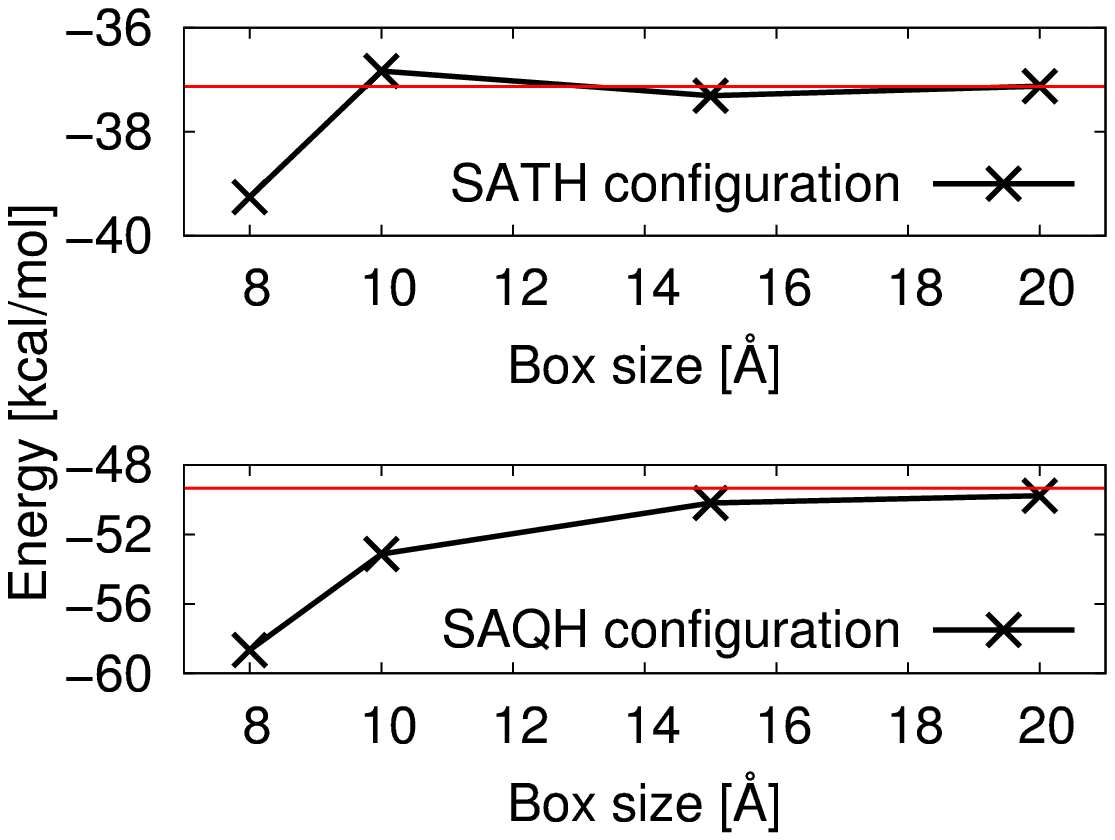}
\par\end{centering}

}
\par\end{centering}

\caption{Geometry optimised configurations for tri- and tetrahydrated (SATH
and SAQH) clusters are shown in (a) and (b) respectively. The labelling
of various hydrogen bonds is referred to in Section \ref{sub:PIMD}.
(c) shows the binding energies of configurations (a) and (b) as a
function of the system box size, converging to values obtained by
\citet{Temelso2012} at the MP2 level.\label{fig:Geometry-optimised-configuration}}
\end{figure}

\begin{figure}
\begin{centering}
\includegraphics[width=1\columnwidth]{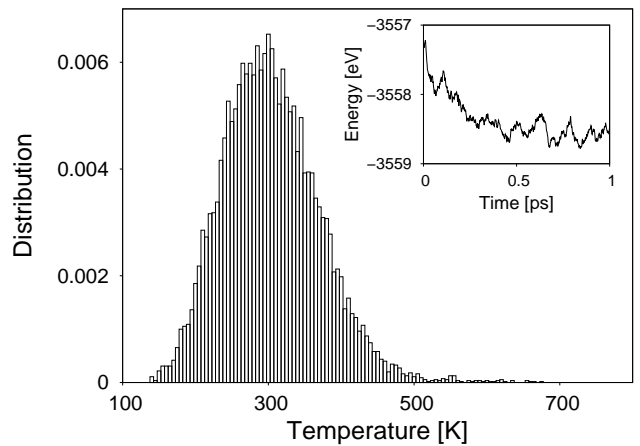}
\par\end{centering}

\caption{\label{fig:TempHistogram}Histogram of cluster kinetic energy (represented
as a temperature) from the equilibrated simulation referred to as
SATH ${\rm 1}$ bead in Table \ref{tab:timeStepData}. The inset shows
the system's energy as a function of time for the first $\mathrm{1\: ps}$
of the same simulation.}
\end{figure}

Molecular dynamics simulations at $\mathrm{300\: K}$ incorporating
both classical nuclear dynamics and PIMD were performed using the
CASTEP \citep{Clark2005} (version ${\rm 5.5}$) code. The standard
on-the-fly ultrasoft pseudopotential provided internally by the CASTEP
code was employed for all calculations. The Perdew-Burke-Ernzerhof
\citep{Perdew1996} (PBE) functional was used with a plane wave basis
set. The PBE functional has been found to perform well for hydrogen
bonded systems\citep{Thanthiriwatte2011,Ireta2004}. A cut off energy
of ${\rm 550\: eV}$ was found to converge the plane wave basis set
sufficiently for all systems studied. A time step of $\mathrm{1\: fs}$
was used for classical (single bead) simulations and a time step of
$\mathrm{0.5\: fs}$ or shorter was used for the PIMD simulations
due to the stiffness of the inter-bead springs. CASTEP utilizes the
Born-Oppenheimer version of ab initio MD and the Langevin thermostat
with a friction constant of ${\rm 0.01\: fs^{-1}}$ was used in all
simulations. The equilibration period was judged by observing when
the running mean energy of the system had relaxed (usually requiring
less than $\mathrm{0.5\: ps}$) and also by monitoring the distribution
of cluster 'temperature' (or kinetic energy in the centre of mass
frame), which ought to be approximately Gaussian \citep{Haile1997}
with a standard deviation ($\sigma$) obeying $\sigma/\left\langle T\right\rangle \sim N^{-1/2}$.
A typical temperature histogram satisfying this requirement is shown
in Figure \ref{fig:TempHistogram}. The inset in Figure \ref{fig:TempHistogram}
shows that the typical relaxation time was in the order of $\mathrm{0.5\: ps}$
relaxation time. Initial configurations of sulphuric acid and water
identified from the literature were constructed under a classical
potential (MMFF94s) using the Avogadro \citep{Hanwell2012} (version
${\rm 1.0.3}$) package. The choices of time step and simulation time
for various cases are given in Table \ref{tab:timeStepData}.

Configurations of ${\rm [H_{2}SO_{4}][H_{2}O]_{n=3-4}}$ were studied
at a target temperature of ${\rm 300\: K}$. PIMD uses a certain number
of beads to approximate the zero-point motion, and cases with $P=1$,
${\rm 4}$, ${\rm 8}$, ${\rm 16}$ and ${\rm 32}$ beads were tested
in this study. The staging transformation\citep{MarkE.Tuckerman2010}
was used for all PIMD simulations. The $P=1$ case represents the
classical limit of the PIMD technique and corresponds to the complete
neglect of zero-point motion. The box size of the system was optimised
against MP${\rm 2}$ level data \citep{Temelso2012} as shown in Figure
\ref{fig:binding energy}(c). The binding energies, at zero temperature,
of the two configurations in Figures \ref{fig:SATH and SAQH configuration(A)}
and \ref{fig:SATH and SAQH configuration(b)} are compared against
MP${\rm 2}$ level data. A box size of ${\rm 15\:\hbox{\AA}}$ was
chosen as a compromise between accuracy and computational demand.

\section{Results\label{sec:3Results}}

\subsection{DFT without zero-point motion\label{sub:DFT}}

\begin{figure}
\noindent \begin{centering}
\subfloat[\label{fig:Config H label}]{\noindent \begin{centering}
\includegraphics[width=0.7\columnwidth]{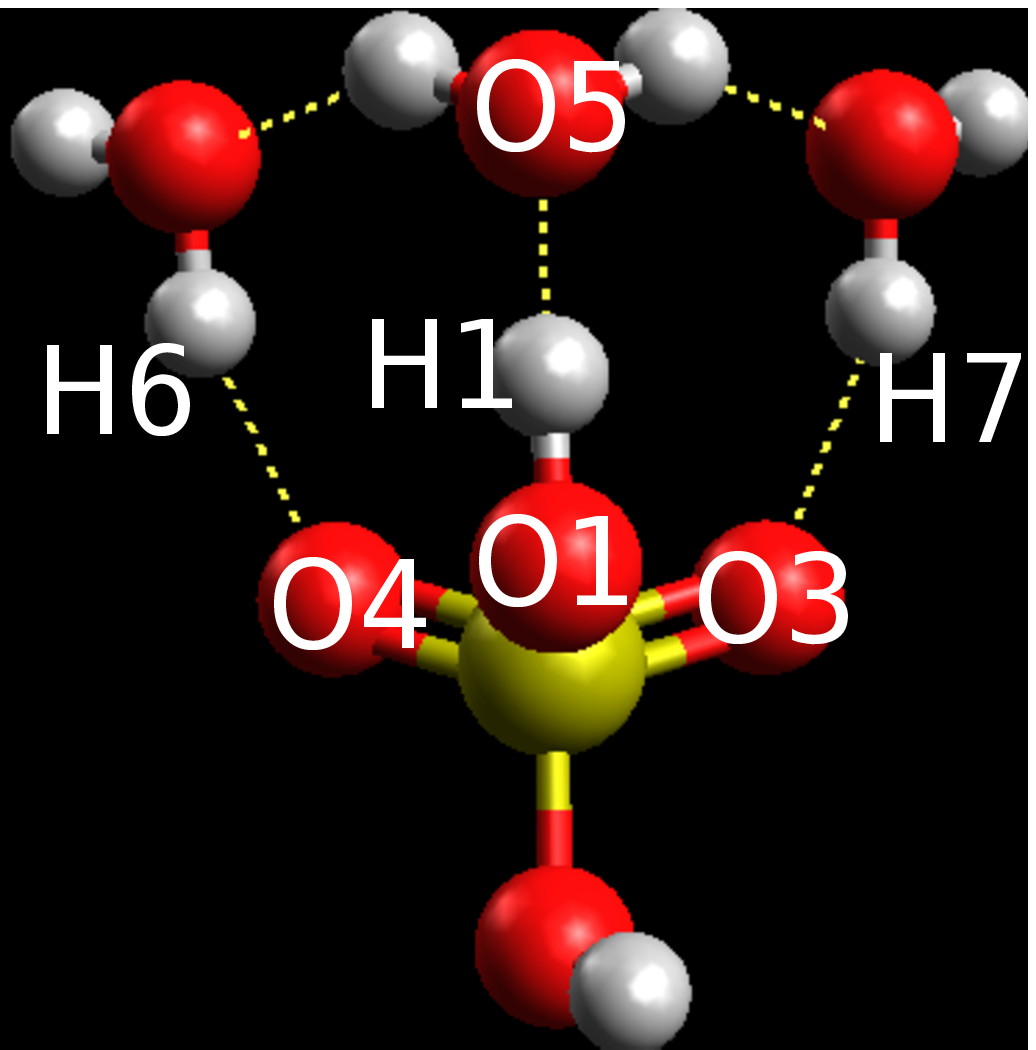}
\par\end{centering}

}
\par\end{centering}

\noindent \begin{centering}
\subfloat[\label{fig:hb hb comparison for DFT}]{\noindent \begin{centering}
\includegraphics[width=0.8\columnwidth]{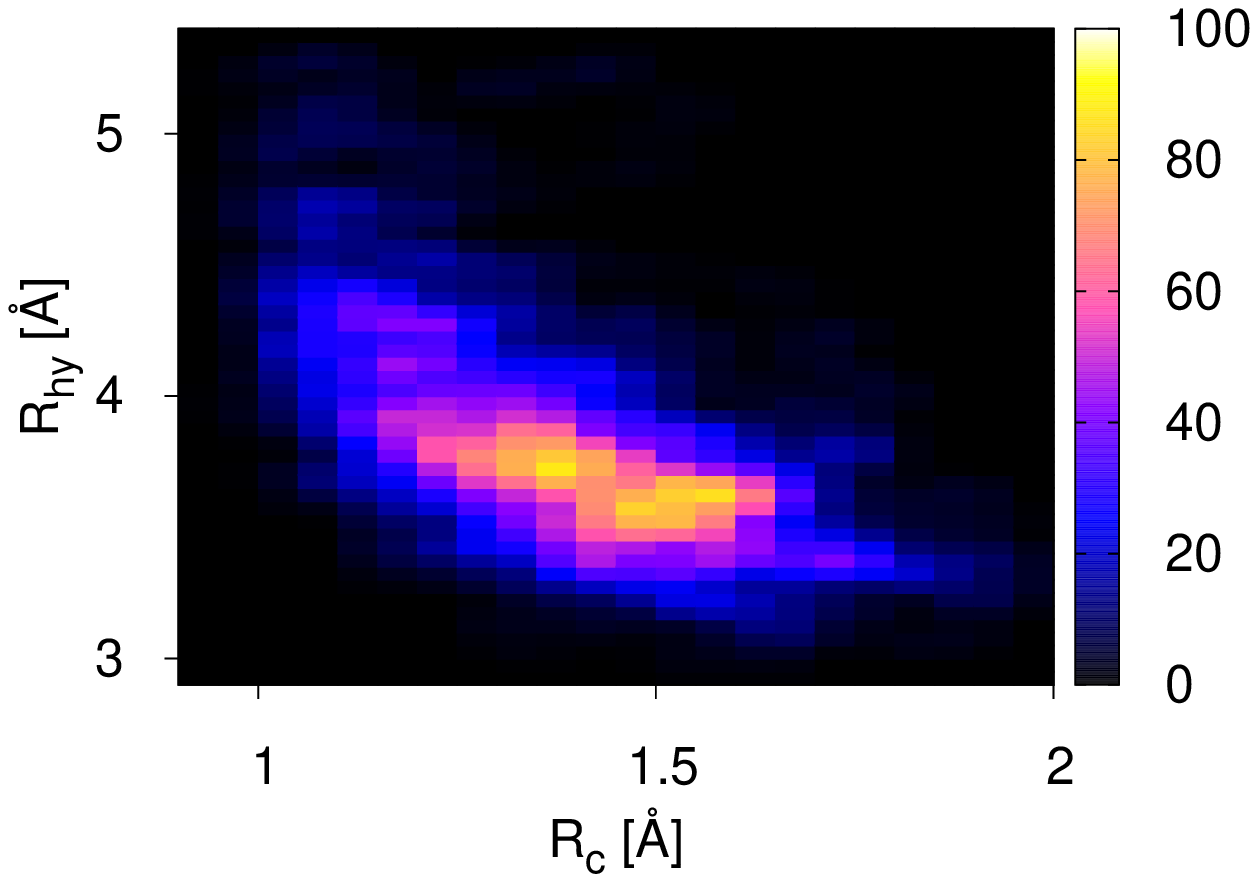}
\par\end{centering}

}
\par\end{centering}

\caption{{\small The configuration denoted config H is shown in (a) with labels
that identify certain O-H pairs. Plot (b) illustrates the probability
density (given in arbitrary units) as a function of two structural
features labelled $R_{{\rm c}}$ (the length of the covalent bond
O1-H1) and $R_{{\rm hy}}$ (the sum of the lengths of prospective
hydrogen bonds O4-H6 and O3-H7), obtained at DFT level, equivalent
to using a single bead in PIMD. The associated potential of mean force
takes the form of a broad, shallow well where the ionisation of the
configuration is correlated with the status of the adjacent hydrogen
bonds, as denoted by $R_{{\rm hy}}$. }}
\end{figure}

Molecular configurations likely to feature a dissociated sulphuric
acid molecule were identified from the literature and investigated.
One such configuration was labelled III-i-1 by \citet{Re1999} and
is illustrated here in Figure \ref{fig:Config H label} and denoted
config H. Our single bead simulations at $\mathrm{300\: K}$ show
that the proton labelled H1 moves with considerable freedom between
oxygens O1 and O5. Furthermore, Figure \ref{fig:hb hb comparison for DFT}
demonstrates an anticorrelation between the length $R_{{\rm c}}$
of the dissociating bond O1-H1 and the sum of the lengths of the neighbouring
hydrogen bonds, labelled O3-H7 and O4-H6 in Figure \ref{fig:Config H label},
and denoted $R_{{\rm hy}}$. The formation of the `ionised' state
due to the switch to the O5-H1 bond (such that the value of $R_{{\rm c}}$
is large) is seen to depend upon the prior existence of both the neighbouring
hydrogen bonds (namely a low value of $R_{{\rm hy}}$). If either
neighbouring hydrogen bond is broken the system remains `neutral'
(with a low value of $R_{{\rm c}}$), which is not surprising since
the configuration is then similar to the SATH structure shown in Figure
\ref{fig:Geometry-optimised-configuration}. This is an important
corollary to conclusions acquired from consideration of geometry optimisation
at $\mathrm{0\: K}$, where config H has been shown to ionise \citep{Re1999}.
At ${\rm 300\: K}$ the behaviour can most certainly not be represented
by harmonic fluctuations about an ionised mean structure and a free
energy based on the rigid-rotor-harmonic-approximation for this configuration
would fail due to significant anharmonic contributions. We shall return
to this system in the next section.

\subsection{PIMD\label{sub:PIMD}}

\begin{figure}
\begin{centering}
\includegraphics[width=1\linewidth]{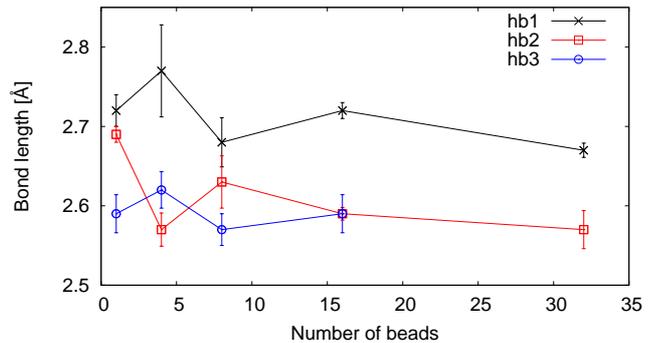}
\par\end{centering}

\centering{}\caption{\label{fig:OOLength}The average oxygen-oxygen separation ${\rm d_{OO}}$
of specific hydrogen bonds as a function of the number of beads used
in the simulation. Labels hb1 and hb2 refer to Figure \ref{fig:SATH and SAQH configuration(A)}
and hb3 is shown in Figure \ref{fig:SATH and SAQH configuration(b)}.
The error bars were determined by the standard blocking procedure
\citep{Flyvbjerg1989,Frenkel2001} and a blocking length of $\mathrm{0.256\: ps}$
was found to give independent sampling. The calculations correspond
to the cases listed in Table \ref{tab:timeStepData}.}
\end{figure}

A PIMD study was performed first for two low energy configurations
(denoted SATH and SAQH) identified in the literature \citep{Re1999,Bandy1998}
and shown in Figures \ref{fig:SATH and SAQH configuration(A)} and
\ref{fig:SATH and SAQH configuration(b)}. It is envisaged that hydrogen
bonds, in particular those associated with the sulphuric acid, would
be the most susceptible to zero-point effects due to the inherent
tendency of sulphuric acid to dissociate. Figure \ref{fig:OOLength}
shows the average oxygen-oxygen distance (${\rm d_{OO}}$) of specific
hydrogen bonds as a function of the number of beads representing atoms
in the system. The bonds labelled hb1 and hb2 in the SATH structure
contract in length by around ${\rm 2-5\%}$ with respect to the outcome
of classical dynamics while the situation for hb3 is less clear. Note
that the longest simulations were performed for the single bead and
16 bead representations of the SATH structure, as indicated in Table
\ref{tab:timeStepData}. For other cases shorter studies were performed
to illustrate the trends, though the accuracy of the results is lower.

\begin{figure}
\begin{centering}
\subfloat[\label{fig:PMFe}]{\begin{centering}
\includegraphics[width=0.35\columnwidth]{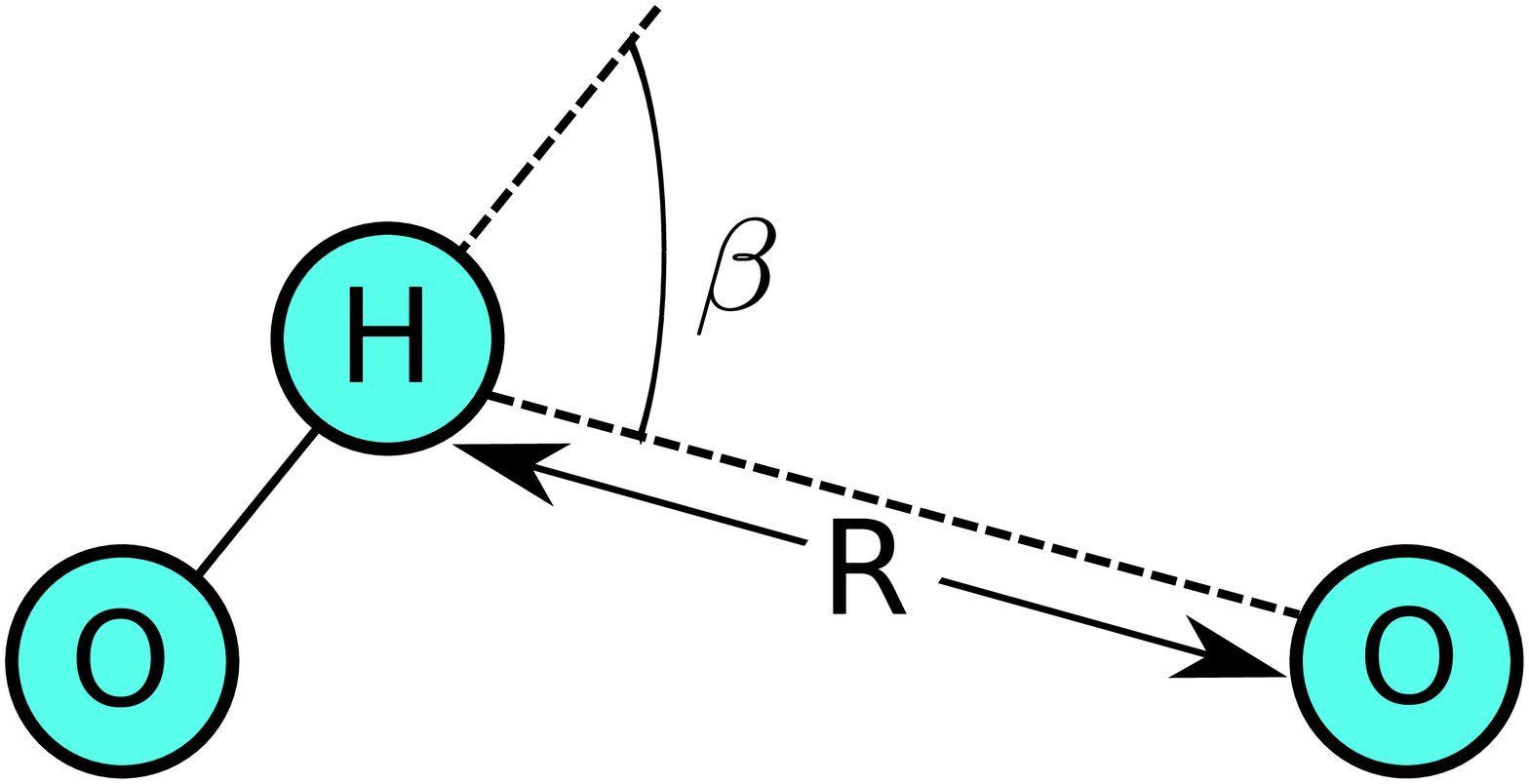}
\par\end{centering}

}
\par\end{centering}

\begin{centering}
\subfloat[\label{fig:PMFb}]{\noindent \centering{}\includegraphics[width=0.85\columnwidth]{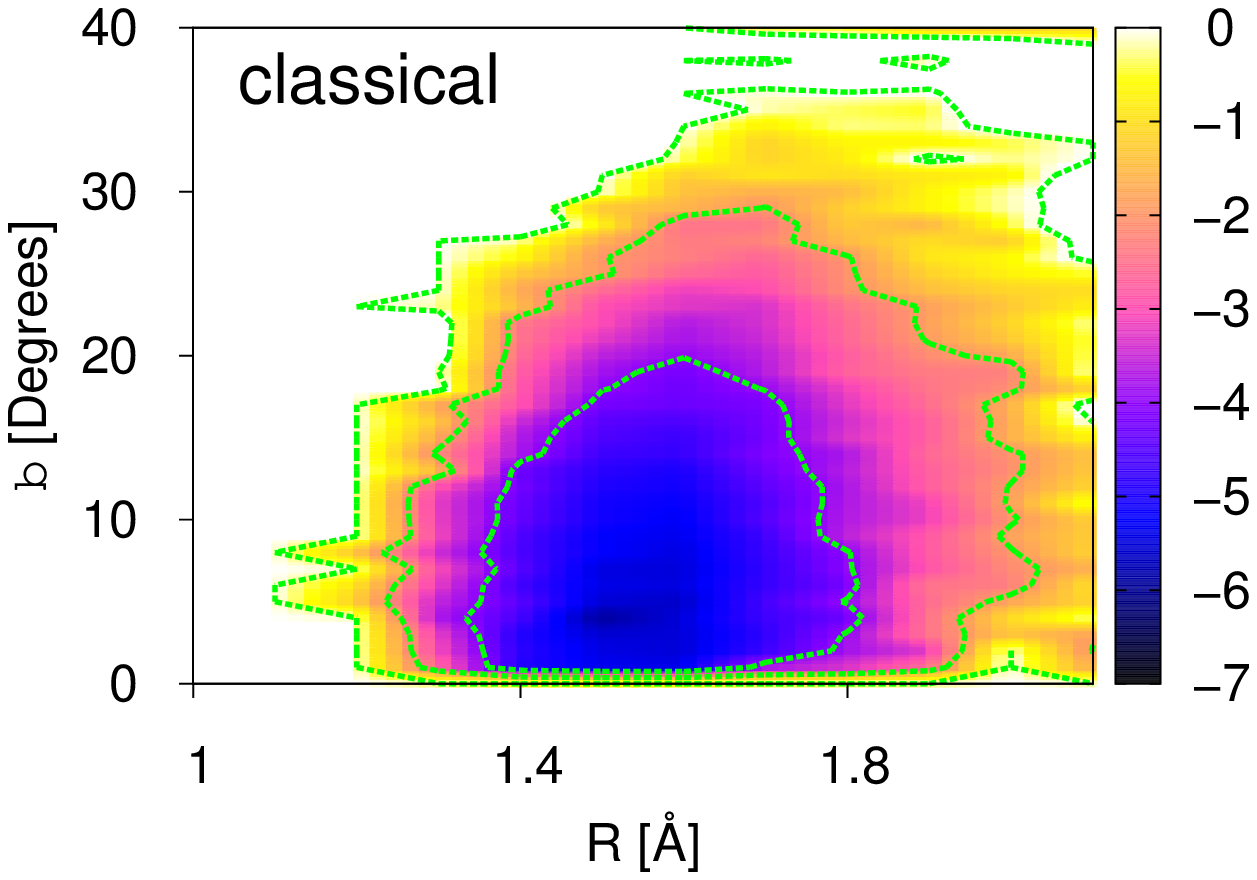}}
\par\end{centering}

\begin{centering}
\subfloat[\label{fig:PMFc}]{\noindent \begin{centering}
\includegraphics[width=0.85\columnwidth]{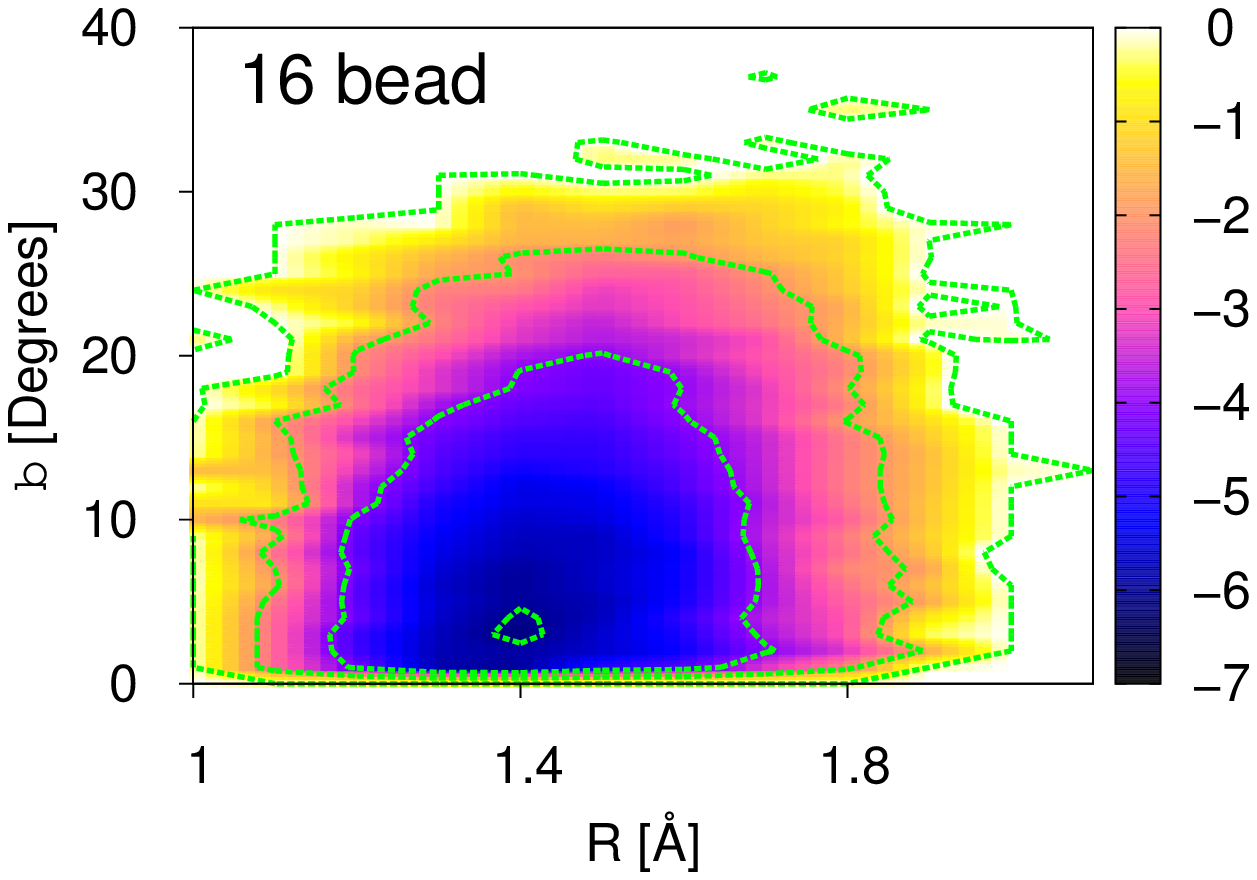}
\par\end{centering}

}
\par\end{centering}

\begin{centering}
\subfloat[\label{fig:PMFd}]{\noindent \begin{centering}
\includegraphics[width=0.85\columnwidth]{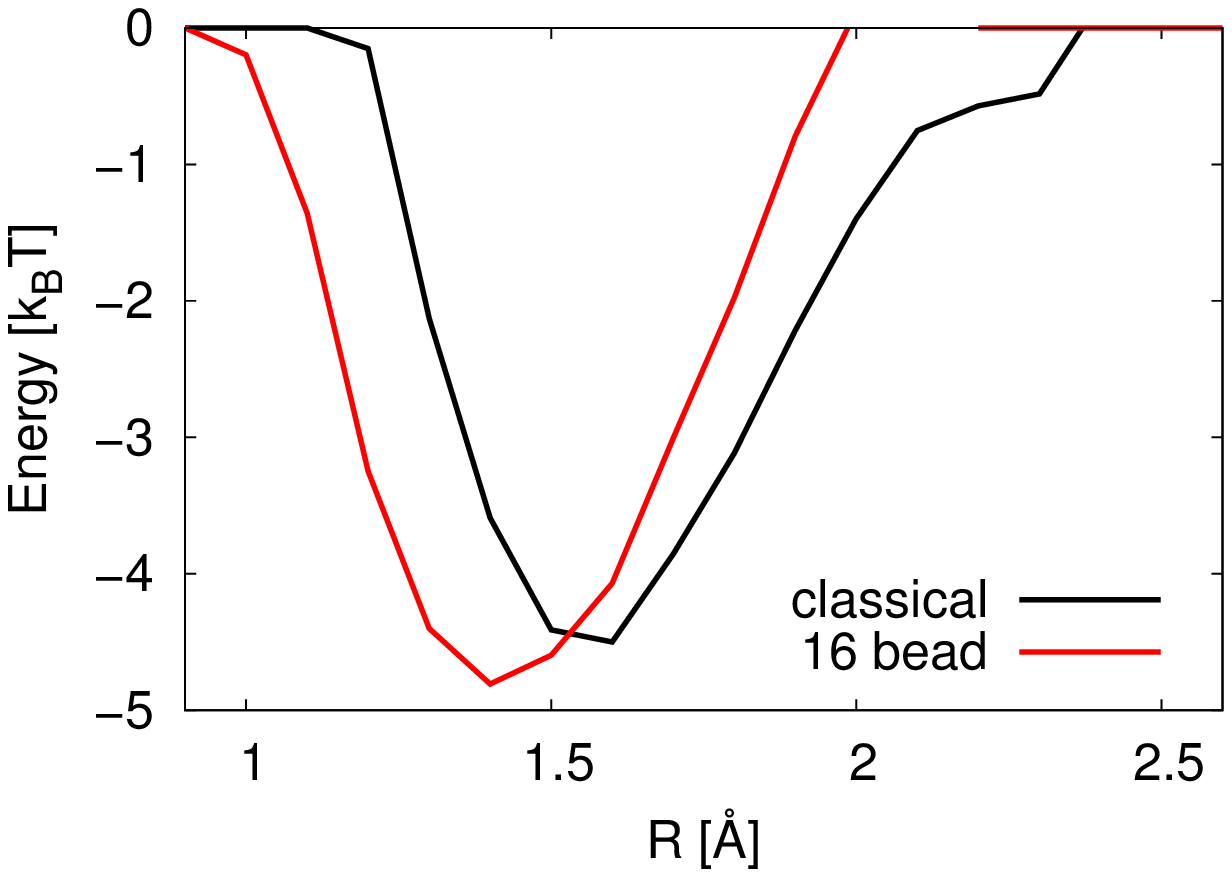}
\par\end{centering}

}
\par\end{centering}

\noindent \caption{\label{fig:PMF} Contour plots of the potential of mean force \foreignlanguage{english}{$W(R,\beta)$}
in units of $k_{B}T$ for the hydrogen in the bond labelled hb2 in
Figure \ref{fig:SATH and SAQH configuration(A)}. The green dashed
lines indicate contour levels of $\mathrm{0}$, $\mathrm{-2}$, $\mathrm{-4}$
and $\mathrm{-6}$. The coordinates for the PMF are defined by sketch
(a) and the method follows the approach described by \citet{Kumar2007}.
Plot (b) shows results from standard DFT molecular dynamics and plot
(c) arises from PIMD using 16 beads. The simulation times are given
in Table \ref{tab:timeStepData}. Plot (d) shows a 1D version of plots
(b) and (c) obtained by integrating over the $\beta$ parameter.}
\end{figure}

Next we examine in detail how the behaviour of the hydrogen atom in
hydrogen bond hb2 is affected by PIMD. This is explored by constructing
a potential of mean force (PMF) for the hydrogen, defined by: 
\[
W(R,\beta)=-k_{B}T\ln g(R,\beta)
\]
where $R$ and $\beta$ are geometric parameters illustrated in Figure
\ref{fig:PMFe} and $g(R,\beta)$ is the proportion of simulation
snapshots with the hydrogen located within the region defined by $R\rightarrow R+dR$
and $\beta\rightarrow\beta+d\beta$ divided by the equivalent proportion
for noninteracting particles. For the PIMD simulations the centroid
of the beads representing the hydrogen atom was used to produce the
PMF. The method is described extensively by \citet{Kumar2007}. Figure
\ref{fig:PMFb} and \ref{fig:PMFc} show the PMFs acquired using classical
MD and PIMD, respectively, for hydrogen bond hb2. 

The PMF plots in Figure \ref{fig:PMF} visualise the differences between
the dynamics of the hb2 bond in Figure \ref{fig:SATH and SAQH configuration(A)}
under classical MD and the PIMD schemes. Such a comparison is limited
by the computationally expensive techniques employed. However it does
offer an insight into the importance of zero-point effects in small
clusters of sulphuric acid and water. The main effect is a shift in
the minimum of the PMF of hydrogen bond length $R$ by about $\mathrm{0.2\:}\textrm{\AA}$
going from the DFT to the PIMD result indicating that the zero-point
motion has a mean configurational influence on this bond. Figure \ref{fig:PMFd}
is a one dimensional version of Figures \ref{fig:PMFb} and \ref{fig:PMFb}
obtained by integrating over the $\beta$ parameter.

\begin{figure}
\begin{centering}
\includegraphics[width=1\columnwidth]{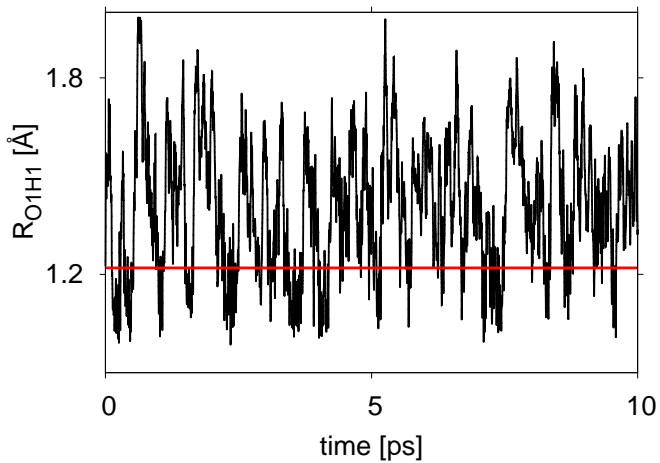}
\par\end{centering}

\caption{\label{fig:BiModal}Plot of the O1-H1 bond length (denoted as $\mathrm{R_{O1H1}}$)
against time in config H in Figure \ref{fig:Config H label} from
the $\mathrm{1}$ bead simulation as detailed in Table \ref{tab:timeStepData}.
The plot clearly illustrates the motion between the neutral and ionised
states. The horizontal line drawn at ${\rm 1.22}\:{\rm \hbox{\AA}}$
provides a simple threshold between covalent and hydrogen bond-like
behaviour of the O1-H1 bond. Under quantum nuclear dynamics the fraction
of time spent above this threshold increases in the order of 10\%.}
\end{figure}

The effects of zero-point motion are clearly rather subtle. To explore
this further, we return to the delicate switching behaviour of the
O1-H1-O5 bonds discussed in section \ref{sub:DFT} and contrast the
classical and quantum nuclear dynamics. Figure \ref{fig:BiModal}
illustrates the motion of the proton between the neutral and ionised
positions, discussed earlier, in terms of the O1-H1 bond length. Which
of the nuclei O1 or O5 was the nearest neighbour to the H1 nucleus
(see Figure \ref{fig:Config H label}) was monitored to quantify this
hopping behaviour. It was found that in the classical case H1 was
closer to O1 for $\mathrm{21.5\%}$ percent of the simulation with
standard error $\sigma_{SE}=3.2\%$ whereas in the 16 bead PIMD simulation
this figure dropped to $\mathrm{14.8\%}$ with $\sigma_{SE}=2.7\%$.
. This property was further investigated by defining a threshold for
the O1-H1 bond length below which the system is considered neutral,
and beyond which it is better described as ionised. We define a ${\rm 1.22\:}\hbox{\AA}$
distance to separate the two regimes, and this is shown as a horizontal
line in Figure \ref{fig:BiModal}. For the classical dynamics, the
percentage of time the system remains neutral according to this criterion
is ${\rm 20.1\%}$ with ${\rm \sigma_{SE}=2.9\%}$. An analysis of
the PIMD simulation with 16 beads yields a corresponding percentage
of neutral residence time of ${\rm 12.5\%}$ with $\sigma_{SE}={\rm 2.4\%}$.
These results are consistent with those determined from the nearest
neighbour criterion. The proportion of time spent in the ionised configuration
rises from $\mathrm{79.1\%}$ to $\mathrm{87.5\%}$. This suggests
that the inclusion of zero-point motion promotes the formation of
the ionised state; quantum uncertainty favours proton transfer.

\section{Conclusions\label{sec:4Conclusion}}

As a consequence of the computational expense of the PIMD technique,
especially when using many beads, the simulations presented are limited
in duration to around $\mathrm{10\: ps}$ for some configurations,
and rather less for others. The statistics on the structural and dynamical
behaviour are therefore preliminary. However, it is possible to extract
some important features from these simulations that correspond to
intuitive expectation, and which can be explored further with more
extensive calculations. 

Our study of small clusters of water and sulphuric acid molecules
leads us to two main conclusions. Firstly, we have demonstrated that
molecular dynamics can reveal features that are not available from
knowledge of the geometry optimised structure at zero temperature.
The prime example of this is the complex behaviour of cluster configuration
III-i-1 identified by \citeauthor{Re1999} \citep{Re1999} and here
denoted config H. This configuration has been regarded as the most
stable ionised configuration for the trihydrated sulphuric acid molecule
\citep{Bandy1998,Re1999,Temelso2012}, but our results indicate that
the structure exhibits both neutral and ionised characteristics at
${\rm 300\: K}$. This conclusion highlights the limitations of the
RRHO approximation\citep{Kathmann2007,Kathmann2007b} for free energy
estimation using a single optimised structure.

Secondly, the inclusion of zero-point motion through PIMD simulations
of the sulphuric acid-water system has been shown to produce small
but clear structural distortion at ${\rm 300\: K}$ in a selected
number of configurations when compared with classical dynamics. The
mean oxygen-oxygen separation of hydrogen bonds hb1 and hb2 in the
structure shown in Figure \ref{fig:SATH and SAQH configuration(A)}
is reduced by ${\rm 2-5\%}$. We observe a mild shortening of the
hb2 hydrogen bond length, shown by constructing potentials of mean
force for the classical and PIMD schemes, as illustrated in Figure
\ref{fig:PMF}. Furthermore, our results indicate that zero-point
motion brings about a greater propensity for proton transfer in the
O1-H1-O5 substructure of the configuration shown in Figure \ref{fig:Config H label}
at ${\rm 300\: K}$.

This conclusion is consistent with the paper by \citet{Li2011b} where
quantum nuclear effects on the hydrogen bond are studied, specifically
Figure 3 in reference \citen{Li2011b} where the OO length is compared
with the length of the projection of the covalent OH bond on the OO
vector. The implication is that the projected covalent bond length
is increased by quantum effects when the hydrogen bond is considered
to be strong, as judged by a shift in vibrational frequency of the
covalent OH bond due to the presence of the hydrogen bond.

Our research supports the view that the zero-point effect is most
significant in configurations where proton transfer is intrinsically
likely. Classical and PIMD simulations of the cluster shown in Figure
\ref{fig:Config H label} have demonstrated frequent proton transfer.
Using an O-H separation of ${\rm 1.22}\:{\rm \hbox{\AA}}$ as a threshold
for distinguishing the ionised from the neutral state, the cluster
is found to remain neutral ${\rm 20.1\%}$ of the time (${\rm \sigma_{SE}=2.9\%}$)
with classical MD and ${\rm 12.5\%}$ ($\sigma_{SE}={\rm 2.4\%}$)
according to PIMD. It is possible to infer that quantum effects have
increased the degree of proton transfer. It is expected that simulations
at lower temperatures would increase the significance of the zero-point
effects, making this an avenue for future research. In addition, since
substances such as ammonia and amines are increasingly thought to
be relevant to atmospheric nucleation \citep{Kurten2008,Kirkby2011},
assessing the importance of zero-point motion in these systems would
also be of interest.

In summary, zero-point motion does affect the structure of small clusters
of sulphuric acid and water, particularly the lengths of hydrogen
bonds. At $\mathrm{300\; K}$, the contribution appears to be most
significant for cases that are intrinsically susceptible to proton
transfer.

\section{Acknowledgements }

We thank Prof. Angelos Michaelides and his group at UCL for practical
advice and helpful discussions and this work benefited from interactions
within the Thomas Young Centre. SMK was supported by the U.S. Department
of Energy, Office of Basic Energy Sciences, Division of Chemical Sciences,
Geosciences, and Biosciences. JLS was supported by the IMPACT scheme
at UCL and by the U.S. Department of Energy, Office of Basic Energy
Sciences, Division of Chemical Sciences, Geosciences, and Biosciences.
We are grateful for use of the UCL Legion High Performance Computing
Facility and the resources of the National Energy Research Scientific
Computing Center (NERSC), which is supported by the U.S. Department
of Energy, Office of Science under Contract No. DE-AC02-05CH11231. 

\bibliographystyle{apsrev4-1}

\begin{thebibliography}{41}%
\makeatletter
\providecommand \@ifxundefined [1]{%
 \@ifx{#1\undefined}
}%
\providecommand \@ifnum [1]{%
 \ifnum #1\expandafter \@firstoftwo
 \else \expandafter \@secondoftwo
 \fi
}%
\providecommand \@ifx [1]{%
 \ifx #1\expandafter \@firstoftwo
 \else \expandafter \@secondoftwo
 \fi
}%
\providecommand \natexlab [1]{#1}%
\providecommand \enquote  [1]{``#1''}%
\providecommand \bibnamefont  [1]{#1}%
\providecommand \bibfnamefont [1]{#1}%
\providecommand \citenamefont [1]{#1}%
\providecommand \href@noop [0]{\@secondoftwo}%
\providecommand \href [0]{\begingroup \@sanitize@url \@href}%
\providecommand \@href[1]{\@@startlink{#1}\@@href}%
\providecommand \@@href[1]{\endgroup#1\@@endlink}%
\providecommand \@sanitize@url [0]{\catcode `\\12\catcode `\$12\catcode
  `\&12\catcode `\#12\catcode `\^12\catcode `\_12\catcode `\%12\relax}%
\providecommand \@@startlink[1]{}%
\providecommand \@@endlink[0]{}%
\providecommand \url  [0]{\begingroup\@sanitize@url \@url }%
\providecommand \@url [1]{\endgroup\@href {#1}{\urlprefix }}%
\providecommand \urlprefix  [0]{URL }%
\providecommand \Eprint [0]{\href }%
\providecommand \doibase [0]{http://dx.doi.org/}%
\providecommand \selectlanguage [0]{\@gobble}%
\providecommand \bibinfo  [0]{\@secondoftwo}%
\providecommand \bibfield  [0]{\@secondoftwo}%
\providecommand \translation [1]{[#1]}%
\providecommand \BibitemOpen [0]{}%
\providecommand \bibitemStop [0]{}%
\providecommand \bibitemNoStop [0]{.\EOS\space}%
\providecommand \EOS [0]{\spacefactor3000\relax}%
\providecommand \BibitemShut  [1]{\csname bibitem#1\endcsname}%
\let\auto@bib@innerbib\@empty
\bibitem [{Roy(2009)}]{RoyalSocietyGeoEngineeringReport2009}%
  \BibitemOpen
  \href@noop {} {\emph {\bibinfo {title} {Geoengineering the climate: science,
  governance and uncertainty}}}\ (\bibinfo  {publisher} {The Royal Society},\
  \bibinfo {year} {2009})\BibitemShut {NoStop}%
\bibitem [{\citenamefont {Zhang}\ \emph {et~al.}(2012)\citenamefont {Zhang},
  \citenamefont {Khalizov}, \citenamefont {Wang}, \citenamefont {Hu},\ and\
  \citenamefont {Xu}}]{Zhang2012}%
  \BibitemOpen
  \bibfield  {author} {\bibinfo {author} {\bibfnamefont {R.}~\bibnamefont
  {Zhang}}, \bibinfo {author} {\bibfnamefont {A.}~\bibnamefont {Khalizov}},
  \bibinfo {author} {\bibfnamefont {L.}~\bibnamefont {Wang}}, \bibinfo {author}
  {\bibfnamefont {M.}~\bibnamefont {Hu}}, \ and\ \bibinfo {author}
  {\bibfnamefont {W.}~\bibnamefont {Xu}},\ }\href {\doibase 10.1021/cr2001756}
  {\bibfield  {journal} {\bibinfo  {journal} {Chem. Rev.}\ }\textbf {\bibinfo
  {volume} {112}},\ \bibinfo {pages} {1957} (\bibinfo {year}
  {2012})}\BibitemShut {NoStop}%
\bibitem [{\citenamefont {Ford}(2004)}]{Ford2004}%
  \BibitemOpen
  \bibfield  {author} {\bibinfo {author} {\bibfnamefont {I.~J.}\ \bibnamefont
  {Ford}},\ }\href {\doibase 10.1243/0954406041474183} {\bibfield  {journal}
  {\bibinfo  {journal} {Proc. Inst. Mech. Eng., Part C: J. Mech. Eng. Sci.}\
  }\textbf {\bibinfo {volume} {218}},\ \bibinfo {pages} {883} (\bibinfo {year}
  {2004})}\BibitemShut {NoStop}%
\bibitem [{\citenamefont {Vehkam\"{a}ki}(2006)}]{Vehkamaki}%
  \BibitemOpen
  \bibfield  {author} {\bibinfo {author} {\bibfnamefont {H.}~\bibnamefont
  {Vehkam\"{a}ki}},\ }\href@noop {} {\emph {\bibinfo {title} {{Classical
  Nucleation Theory in Multicomponent Systems}}}}\ (\bibinfo  {publisher}
  {Springer},\ \bibinfo {year} {2006})\BibitemShut {NoStop}%
\bibitem [{\citenamefont {Napari}\ \emph {et~al.}(2010)\citenamefont {Napari},
  \citenamefont {Julin},\ and\ \citenamefont {Vehkam\"{a}ki}}]{Napari2010}%
  \BibitemOpen
  \bibfield  {author} {\bibinfo {author} {\bibfnamefont {I.}~\bibnamefont
  {Napari}}, \bibinfo {author} {\bibfnamefont {J.}~\bibnamefont {Julin}}, \
  and\ \bibinfo {author} {\bibfnamefont {H.}~\bibnamefont {Vehkam\"{a}ki}},\
  }\href {\doibase 10.1063/1.3502643} {\bibfield  {journal} {\bibinfo
  {journal} {J. Chem. Phys.}\ }\textbf {\bibinfo {volume} {133}},\ \bibinfo
  {pages} {154503} (\bibinfo {year} {2010})}\BibitemShut {NoStop}%
\bibitem [{\citenamefont {Re}\ \emph {et~al.}(1999)\citenamefont {Re},
  \citenamefont {Osamura},\ and\ \citenamefont {Morokuma}}]{Re1999}%
  \BibitemOpen
  \bibfield  {author} {\bibinfo {author} {\bibfnamefont {S.}~\bibnamefont
  {Re}}, \bibinfo {author} {\bibfnamefont {Y.}~\bibnamefont {Osamura}}, \ and\
  \bibinfo {author} {\bibfnamefont {K.}~\bibnamefont {Morokuma}},\ }\href
  {\doibase 10.1021/jp984759x} {\bibfield  {journal} {\bibinfo  {journal} {J.
  Phys. Chem. A}\ }\textbf {\bibinfo {volume} {103}},\ \bibinfo {pages} {3535}
  (\bibinfo {year} {1999})}\BibitemShut {NoStop}%
\bibitem [{\citenamefont {Bandy}\ and\ \citenamefont
  {Ianni}(1998)}]{Bandy1998}%
  \BibitemOpen
  \bibfield  {author} {\bibinfo {author} {\bibfnamefont {A.~R.}\ \bibnamefont
  {Bandy}}\ and\ \bibinfo {author} {\bibfnamefont {J.~C.}\ \bibnamefont
  {Ianni}},\ }\href {\doibase 10.1021/jp980270s} {\bibfield  {journal}
  {\bibinfo  {journal} {J. Phys. Chem. A}\ }\textbf {\bibinfo {volume} {102}},\
  \bibinfo {pages} {6533} (\bibinfo {year} {1998})}\BibitemShut {NoStop}%
\bibitem [{\citenamefont {Ianni}\ and\ \citenamefont
  {Bandy}(1999)}]{Ianni2000}%
  \BibitemOpen
  \bibfield  {author} {\bibinfo {author} {\bibfnamefont {J.~C.}\ \bibnamefont
  {Ianni}}\ and\ \bibinfo {author} {\bibfnamefont {A.~R.}\ \bibnamefont
  {Bandy}},\ }\href {\doibase 10.1016/S0166-1280(99)00182-7} {\bibfield
  {journal} {\bibinfo  {journal} {J. Mol. Struc-THEOCHEM}\ }\textbf {\bibinfo
  {volume} {497}},\ \bibinfo {pages} {19} (\bibinfo {year} {1999})}\BibitemShut
  {NoStop}%
\bibitem [{\citenamefont {Arstila}\ \emph {et~al.}(1998)\citenamefont
  {Arstila}, \citenamefont {Laasonen},\ and\ \citenamefont
  {Laaksonen}}]{Arstila1998}%
  \BibitemOpen
  \bibfield  {author} {\bibinfo {author} {\bibfnamefont {H.}~\bibnamefont
  {Arstila}}, \bibinfo {author} {\bibfnamefont {K.}~\bibnamefont {Laasonen}}, \
  and\ \bibinfo {author} {\bibfnamefont {A.}~\bibnamefont {Laaksonen}},\
  }\href{\doibase 10.1063/1.475496}
  {\bibfield  {journal} {\bibinfo  {journal} {J. Phys. Chem. A}\
  }\textbf {\bibinfo {volume} {108}},\ \bibinfo {pages} {1031} (\bibinfo {year}
  {1998})}\BibitemShut {NoStop}%
\bibitem [{\citenamefont {Larson}\ \emph {et~al.}(2000)\citenamefont {Larson},
  \citenamefont {Kuno},\ and\ \citenamefont {Tao}}]{Larson2000}%
  \BibitemOpen
  \bibfield  {author} {\bibinfo {author} {\bibfnamefont {L.~J.}\ \bibnamefont
  {Larson}}, \bibinfo {author} {\bibfnamefont {M.}~\bibnamefont {Kuno}}, \ and\
  \bibinfo {author} {\bibfnamefont {F.-M.}\ \bibnamefont {Tao}},\ }\href
  {\doibase 10.1063/1.481532} {\bibfield  {journal} {\bibinfo  {journal} {J.
  Chem. Phys.}\ }\textbf {\bibinfo {volume} {112}},\ \bibinfo {pages} {8830}
  (\bibinfo {year} {2000})}\BibitemShut {NoStop}%
\bibitem [{\citenamefont {Ding}\ \emph {et~al.}(2003)\citenamefont {Ding},
  \citenamefont {Laasonen},\ and\ \citenamefont {Laaksonen}}]{Ding2003}%
  \BibitemOpen
  \bibfield  {author} {\bibinfo {author} {\bibfnamefont {C.-G.}\ \bibnamefont
  {Ding}}, \bibinfo {author} {\bibfnamefont {K.}~\bibnamefont {Laasonen}}, \
  and\ \bibinfo {author} {\bibfnamefont {A.}~\bibnamefont {Laaksonen}},\ }\href
  {\doibase 10.1021/jp022575j} {\bibfield  {journal} {\bibinfo  {journal} {J.
  Phys. Chem. A}\ }\textbf {\bibinfo {volume} {107}},\ \bibinfo {pages} {8648}
  (\bibinfo {year} {2003})}\BibitemShut {NoStop}%
\bibitem [{\citenamefont {{A. Natsheh}}\ \emph {et~al.}(2004)\citenamefont {{A.
  Natsheh}}, \citenamefont {Nadykto}, \citenamefont {Mikkelsen}, \citenamefont
  {Yu},\ and\ \citenamefont {Ruuskanen}}]{AlNatsheh2004}%
  \BibitemOpen
  \bibfield  {author} {\bibinfo {author} {\bibfnamefont {A.}~\bibnamefont {{A.
  Natsheh}}}, \bibinfo {author} {\bibfnamefont {A.~B.}\ \bibnamefont
  {Nadykto}}, \bibinfo {author} {\bibfnamefont {K.~V.}\ \bibnamefont
  {Mikkelsen}}, \bibinfo {author} {\bibfnamefont {F.}~\bibnamefont {Yu}}, \
  and\ \bibinfo {author} {\bibfnamefont {J.}~\bibnamefont {Ruuskanen}},\ }\href
  {\doibase 10.1021/jp048858o} {\bibfield  {journal} {\bibinfo  {journal} {J.
  Phys. Chem. A}\ }\textbf {\bibinfo {volume} {108}},\ \bibinfo {pages} {8914}
  (\bibinfo {year} {2004})}\BibitemShut {NoStop}%
\bibitem [{\citenamefont {Arrouvel}\ \emph {et~al.}(2005)\citenamefont
  {Arrouvel}, \citenamefont {Viossat},\ and\ \citenamefont
  {Minot}}]{Arrouvel2005}%
  \BibitemOpen
  \bibfield  {author} {\bibinfo {author} {\bibfnamefont {C.}~\bibnamefont
  {Arrouvel}}, \bibinfo {author} {\bibfnamefont {V.}~\bibnamefont {Viossat}}, \
  and\ \bibinfo {author} {\bibfnamefont {C.}~\bibnamefont {Minot}},\ }\href
  {\doibase 10.1016/j.theochem.2004.12.032} {\bibfield  {journal} {\bibinfo
  {journal} {J. Mol. Struc-THEOCHEM}\ }\textbf {\bibinfo {volume} {718}},\
  \bibinfo {pages} {71} (\bibinfo {year} {2005})}\BibitemShut {NoStop}%
\bibitem [{\citenamefont {Ding}\ and\ \citenamefont
  {Laasonen}(2004)}]{Ding2004}%
  \BibitemOpen
  \bibfield  {author} {\bibinfo {author} {\bibfnamefont {C.-G.}\ \bibnamefont
  {Ding}}\ and\ \bibinfo {author} {\bibfnamefont {K.}~\bibnamefont
  {Laasonen}},\ }\href {\doibase 10.1016/j.cplett.2004.02.112} {\bibfield
  {journal} {\bibinfo  {journal} {Chem. Phys. Lett.}\ }\textbf {\bibinfo
  {volume} {390}},\ \bibinfo {pages} {307} (\bibinfo {year}
  {2004})}\BibitemShut {NoStop}%
\bibitem [{\citenamefont {Nadykto}\ \emph {et~al.}(2008)\citenamefont
  {Nadykto}, \citenamefont {Yu},\ and\ \citenamefont {Herb}}]{Nadykto2008}%
  \BibitemOpen
  \bibfield  {author} {\bibinfo {author} {\bibfnamefont {A.~B.}\ \bibnamefont
  {Nadykto}}, \bibinfo {author} {\bibfnamefont {F.}~\bibnamefont {Yu}}, \ and\
  \bibinfo {author} {\bibfnamefont {J.}~\bibnamefont {Herb}},\ }\href {\doibase
  10.1039/b807415a} {\bibfield  {journal} {\bibinfo  {journal} {Phys. Chem.
  Chem. Phys.}\ }\textbf {\bibinfo {volume} {10}},\ \bibinfo {pages} {7073}
  (\bibinfo {year} {2008})}\BibitemShut {NoStop}%
\bibitem [{\citenamefont {Kurt\'{e}n}\ \emph {et~al.}(2009)\citenamefont
  {Kurt\'{e}n}, \citenamefont {Ortega},\ and\ \citenamefont
  {Vehkam\"{a}ki}}]{Kurten2009}%
  \BibitemOpen
  \bibfield  {author} {\bibinfo {author} {\bibfnamefont {T.}~\bibnamefont
  {Kurt\'{e}n}}, \bibinfo {author} {\bibfnamefont {I.~K.}\ \bibnamefont
  {Ortega}}, \ and\ \bibinfo {author} {\bibfnamefont {H.}~\bibnamefont
  {Vehkam\"{a}ki}},\ }\href {\doibase 10.1016/j.theochem.2009.01.024}
  {\bibfield  {journal} {\bibinfo  {journal} {J. Mol. Struc-THEOCHEM}\ }\textbf
  {\bibinfo {volume} {901}},\ \bibinfo {pages} {169} (\bibinfo {year}
  {2009})}\BibitemShut {NoStop}%
\bibitem [{\citenamefont {Temelso}\ \emph {et~al.}(2012)\citenamefont
  {Temelso}, \citenamefont {Morrell}, \citenamefont {Shields}, \citenamefont
  {Allodi}, \citenamefont {Wood}, \citenamefont {Kirschner}, \citenamefont
  {Castonguay}, \citenamefont {Archer},\ and\ \citenamefont
  {Shields}}]{Temelso2012}%
  \BibitemOpen
  \bibfield  {author} {\bibinfo {author} {\bibfnamefont {B.}~\bibnamefont
  {Temelso}}, \bibinfo {author} {\bibfnamefont {T.~E.}\ \bibnamefont
  {Morrell}}, \bibinfo {author} {\bibfnamefont {R.~M.}\ \bibnamefont
  {Shields}}, \bibinfo {author} {\bibfnamefont {M.~A.}\ \bibnamefont {Allodi}},
  \bibinfo {author} {\bibfnamefont {E.~K.}\ \bibnamefont {Wood}}, \bibinfo
  {author} {\bibfnamefont {K.~N.}\ \bibnamefont {Kirschner}}, \bibinfo {author}
  {\bibfnamefont {T.~C.}\ \bibnamefont {Castonguay}}, \bibinfo {author}
  {\bibfnamefont {K.~A.}\ \bibnamefont {Archer}}, \ and\ \bibinfo {author}
  {\bibfnamefont {G.~C.}\ \bibnamefont {Shields}},\ }\href {\doibase
  10.1021/jp2119026} {\bibfield  {journal} {\bibinfo  {journal} {J. Phys. Chem.
  A}\ }\textbf {\bibinfo {volume} {116}},\ \bibinfo {pages} {2209} (\bibinfo
  {year} {2012})}\BibitemShut {NoStop}%
\bibitem [{\citenamefont {Martin}(2004)}]{RichardMMartin2004}%
  \BibitemOpen
  \bibfield  {author} {\bibinfo {author} {\bibfnamefont {R.~M.}\ \bibnamefont
  {Martin}},\ }\href@noop {} {\emph {\bibinfo {title} {{Electronic structure:
  Basic Theory and practical methods}}}}\ (\bibinfo  {publisher} {Cambridge
  University Press},\ \bibinfo {year} {2004})\BibitemShut {NoStop}%
\bibitem [{\citenamefont {Choe}\ \emph {et~al.}(2007)\citenamefont {Choe},
  \citenamefont {Tsuchida},\ and\ \citenamefont {Ikeshoji}}]{Choe2007}%
  \BibitemOpen
  \bibfield  {author} {\bibinfo {author} {\bibfnamefont {Y.-K.}\ \bibnamefont
  {Choe}}, \bibinfo {author} {\bibfnamefont {E.}~\bibnamefont {Tsuchida}}, \
  and\ \bibinfo {author} {\bibfnamefont {T.}~\bibnamefont {Ikeshoji}},\ }\href
  {\doibase 10.1063/1.2718526} {\bibfield  {journal} {\bibinfo  {journal} {J.
  Chem. Phys.}\ }\textbf {\bibinfo {volume} {126}},\ \bibinfo {pages} {154510}
  (\bibinfo {year} {2007})}\BibitemShut {NoStop}%
\bibitem [{\citenamefont {Anderson}\ \emph {et~al.}(2008)\citenamefont
  {Anderson}, \citenamefont {Siepmann}, \citenamefont {McMurry},\ and\
  \citenamefont {VandeVondele}}]{Anderson2008}%
  \BibitemOpen
  \bibfield  {author} {\bibinfo {author} {\bibfnamefont {K.~E.}\ \bibnamefont
  {Anderson}}, \bibinfo {author} {\bibfnamefont {J.~I.}\ \bibnamefont
  {Siepmann}}, \bibinfo {author} {\bibfnamefont {P.~H.}\ \bibnamefont
  {McMurry}}, \ and\ \bibinfo {author} {\bibfnamefont {J.}~\bibnamefont
  {VandeVondele}},\ }\href {\doibase 10.1021/ja8019774} {\bibfield  {journal}
  {\bibinfo  {journal} {J. Am. Chem. Soc.}\ }\textbf {\bibinfo {volume}
  {130}},\ \bibinfo {pages} {14144} (\bibinfo {year} {2008})}\BibitemShut
  {NoStop}%
\bibitem [{\citenamefont {Hammerich}\ \emph {et~al.}(2008)\citenamefont
  {Hammerich}, \citenamefont {Buch},\ and\ \citenamefont
  {Mohamed}}]{Hammerich2008}%
  \BibitemOpen
  \bibfield  {author} {\bibinfo {author} {\bibfnamefont {A.}~\bibnamefont
  {Hammerich}}, \bibinfo {author} {\bibfnamefont {V.}~\bibnamefont {Buch}}, \
  and\ \bibinfo {author} {\bibfnamefont {F.}~\bibnamefont {Mohamed}},\ }\href
  {\doibase 10.1016/j.cplett.2008.06.053} {\bibfield  {journal} {\bibinfo
  {journal} {Chem. Phys. Lett.}\ }\textbf {\bibinfo {volume} {460}},\ \bibinfo
  {pages} {423} (\bibinfo {year} {2008})}\BibitemShut {NoStop}%
\bibitem [{\citenamefont {Tuckerman}(2010)}]{MarkE.Tuckerman2010}%
  \BibitemOpen
  \bibfield  {author} {\bibinfo {author} {\bibfnamefont {M.~E.}\ \bibnamefont
  {Tuckerman}},\ }\href@noop {} {\emph {\bibinfo {title} {{Statistical
  Mechanics: Theory and Molecular Simulations}}}}\ (\bibinfo  {publisher}
  {Oxford University Press},\ \bibinfo {year} {2010})\BibitemShut {NoStop}%
\bibitem [{\citenamefont {Li}\ \emph {et~al.}(2011)\citenamefont {Li},
  \citenamefont {Walker},\ and\ \citenamefont {Michaelides}}]{Li2011b}%
  \BibitemOpen
  \bibfield  {author} {\bibinfo {author} {\bibfnamefont {X.-Z.}\ \bibnamefont
  {Li}}, \bibinfo {author} {\bibfnamefont {B.}~\bibnamefont {Walker}}, \ and\
  \bibinfo {author} {\bibfnamefont {A.}~\bibnamefont {Michaelides}},\ }\href
  {\doibase 10.1073/pnas.1016653108} {\bibfield  {journal} {\bibinfo  {journal}
  {Proceedings of the National Academy of Sciences}\ }\textbf {\bibinfo
  {volume} {108}},\ \bibinfo {pages} {6369} (\bibinfo {year}
  {2011})}\BibitemShut {NoStop}%
\bibitem [{\citenamefont {Walker}\ and\ \citenamefont
  {Michaelides}(2010)}]{Walker2010}%
  \BibitemOpen
  \bibfield  {author} {\bibinfo {author} {\bibfnamefont {B.}~\bibnamefont
  {Walker}}\ and\ \bibinfo {author} {\bibfnamefont {A.}~\bibnamefont
  {Michaelides}},\ }\href {\doibase 10.1063/1.3505038} {\bibfield  {journal}
  {\bibinfo  {journal} {J. Chem. Phys.}\ }\textbf {\bibinfo {volume} {133}},\
  \bibinfo {pages} {174306} (\bibinfo {year} {2010})}\BibitemShut {NoStop}%
\bibitem [{\citenamefont {Stewart}(2007)}]{Stewart2007}%
  \BibitemOpen
  \bibfield  {author} {\bibinfo {author} {\bibfnamefont {J.~J.~P.}\
  \bibnamefont {Stewart}},\ }\href {\doibase 10.1007/s00894-007-0233-4}
  {\bibfield  {journal} {\bibinfo  {journal} {J. Mol. Model.}\ }\textbf
  {\bibinfo {volume} {13}},\ \bibinfo {pages} {1173} (\bibinfo {year}
  {2007})}\BibitemShut {NoStop}%
\bibitem [{\citenamefont {Kakizaki}\ \emph {et~al.}(2009)\citenamefont
  {Kakizaki}, \citenamefont {Motegi}, \citenamefont {Yoshikawa}, \citenamefont
  {Takayanagi}, \citenamefont {Shiga},\ and\ \citenamefont
  {Tachikawa}}]{Kakizaki2009}%
  \BibitemOpen
  \bibfield  {author} {\bibinfo {author} {\bibfnamefont {A.}~\bibnamefont
  {Kakizaki}}, \bibinfo {author} {\bibfnamefont {H.}~\bibnamefont {Motegi}},
  \bibinfo {author} {\bibfnamefont {T.}~\bibnamefont {Yoshikawa}}, \bibinfo
  {author} {\bibfnamefont {T.}~\bibnamefont {Takayanagi}}, \bibinfo {author}
  {\bibfnamefont {M.}~\bibnamefont {Shiga}}, \ and\ \bibinfo {author}
  {\bibfnamefont {M.}~\bibnamefont {Tachikawa}},\ }\href {\doibase
  10.1016/j.theochem.2009.01.022} {\bibfield  {journal} {\bibinfo  {journal}
  {J. Mol. Struc-THEOCHEM}\ }\textbf {\bibinfo {volume} {901}},\ \bibinfo
  {pages} {1} (\bibinfo {year} {2009})}\BibitemShut {NoStop}%
\bibitem [{\citenamefont {Sugawara}\ \emph {et~al.}(2011)\citenamefont
  {Sugawara}, \citenamefont {Yoshikawa}, \citenamefont {Takayanagi},
  \citenamefont {Shiga},\ and\ \citenamefont {Tachikawa}}]{Sugawara2011}%
  \BibitemOpen
  \bibfield  {author} {\bibinfo {author} {\bibfnamefont {S.}~\bibnamefont
  {Sugawara}}, \bibinfo {author} {\bibfnamefont {T.}~\bibnamefont {Yoshikawa}},
  \bibinfo {author} {\bibfnamefont {T.}~\bibnamefont {Takayanagi}}, \bibinfo
  {author} {\bibfnamefont {M.}~\bibnamefont {Shiga}}, \ and\ \bibinfo {author}
  {\bibfnamefont {M.}~\bibnamefont {Tachikawa}},\ }\href {\doibase
  10.1021/jp202380h} {\bibfield  {journal} {\bibinfo  {journal} {J. Phys. Chem.
  A}\ }\textbf {\bibinfo {volume} {115}},\ \bibinfo {pages} {11486} (\bibinfo
  {year} {2011})}\BibitemShut {NoStop}%
\bibitem [{\citenamefont {Feynman}\ and\ \citenamefont
  {Hibbs}(2010)}]{Feynman2010}%
  \BibitemOpen
  \bibfield  {author} {\bibinfo {author} {\bibfnamefont {R.~P.}\ \bibnamefont
  {Feynman}}\ and\ \bibinfo {author} {\bibfnamefont {A.~R.}\ \bibnamefont
  {Hibbs}},\ }\href
  {http://www.amazon.com/Quantum-Mechanics-Path-Integrals-Emended/dp/0486477223}
  {\emph {\bibinfo {title} {{Quantum Mechanics and Path Integrals: Emended
  Edition (Dover Books on Physics)}}}}\ (\bibinfo  {publisher} {Dover
  Publications},\ \bibinfo {year} {2010})\BibitemShut {NoStop}%
\bibitem [{\citenamefont {Clark}\ \emph {et~al.}(2005)\citenamefont {Clark},
  \citenamefont {Segall},\ and\ \citenamefont {Pickard}}]{Clark2005}%
  \BibitemOpen
  \bibfield  {author} {\bibinfo {author} {\bibfnamefont {S.~J.}\ \bibnamefont
  {Clark}}, \bibinfo {author} {\bibfnamefont {M.~D.}\ \bibnamefont {Segall}}, \
  and\ \bibinfo {author} {\bibfnamefont {C.~J.}\ \bibnamefont {Pickard}},\
  }\href {\doibase 10.1524/zkri.220.5.567.65075} {\bibfield  {journal}
  {\bibinfo  {journal} {Z. Kristallogr.}\ }\textbf {\bibinfo {volume} {220}},\
  \bibinfo {pages} {567} (\bibinfo {year} {2005})}\BibitemShut {NoStop}%
\bibitem [{\citenamefont {Perdew}\ \emph {et~al.}(1996)\citenamefont {Perdew},
  \citenamefont {Burke},\ and\ \citenamefont {Ernzerhof}}]{Perdew1996}%
  \BibitemOpen
  \bibfield  {author} {\bibinfo {author} {\bibfnamefont {J.}~\bibnamefont
  {Perdew}}, \bibinfo {author} {\bibfnamefont {K.}~\bibnamefont {Burke}}, \
  and\ \bibinfo {author} {\bibfnamefont {M.}~\bibnamefont {Ernzerhof}},\ }\href
  {http://www.ncbi.nlm.nih.gov/pubmed/10062328} {\bibfield  {journal} {\bibinfo
   {journal} {Phys. Rev. Lett.}\ }\textbf {\bibinfo {volume} {77}},\ \bibinfo
  {pages} {3865} (\bibinfo {year} {1996})}\BibitemShut {NoStop}%
\bibitem [{\citenamefont {Thanthiriwatte}\ \emph {et~al.}(2011)\citenamefont
  {Thanthiriwatte}, \citenamefont {Hohenstein}, \citenamefont {Burns},\ and\
  \citenamefont {Sherrill}}]{Thanthiriwatte2011}%
  \BibitemOpen
  \bibfield  {author} {\bibinfo {author} {\bibfnamefont {K.~S.}\ \bibnamefont
  {Thanthiriwatte}}, \bibinfo {author} {\bibfnamefont {E.~G.}\ \bibnamefont
  {Hohenstein}}, \bibinfo {author} {\bibfnamefont {L.~A.}\ \bibnamefont
  {Burns}}, \ and\ \bibinfo {author} {\bibfnamefont {C.~D.}\ \bibnamefont
  {Sherrill}},\ }\href {\doibase 10.1021/ct100469b} {\bibfield  {journal}
  {\bibinfo  {journal} {J. Chem. Theory Compt.}\ }\textbf {\bibinfo {volume}
  {7}},\ \bibinfo {pages} {88} (\bibinfo {year} {2011})}\BibitemShut {NoStop}%
\bibitem [{\citenamefont {Ireta}\ \emph {et~al.}(2004)\citenamefont {Ireta},
  \citenamefont {Neugebauer},\ and\ \citenamefont {Scheffler}}]{Ireta2004}%
  \BibitemOpen
  \bibfield  {author} {\bibinfo {author} {\bibfnamefont {J.}~\bibnamefont
  {Ireta}}, \bibinfo {author} {\bibfnamefont {J.}~\bibnamefont {Neugebauer}}, \
  and\ \bibinfo {author} {\bibfnamefont {M.}~\bibnamefont {Scheffler}},\ }\href
  {\doibase 10.1021/jp0377073} {\bibfield  {journal} {\bibinfo  {journal} {J.
  Chem. Phys. A}\ }\textbf {\bibinfo {volume} {108}},\ \bibinfo {pages} {5692}
  (\bibinfo {year} {2004})}\BibitemShut {NoStop}%
\bibitem [{\citenamefont {Haile}(1997)}]{Haile1997}%
  \BibitemOpen
  \bibfield  {author} {\bibinfo {author} {\bibfnamefont {J.~M.}\ \bibnamefont
  {Haile}},\ }\href
  {http://www.amazon.com/Molecular-Dynamics-Simulation-Elementary-Professional/dp/047118439X}
  {\emph {\bibinfo {title} {{Molecular Dynamics Simulation: Elementary Methods
  (Wiley Professional)}}}}\ (\bibinfo  {publisher} {Wiley-Interscience},\
  \bibinfo {year} {1997})\ p.\ \bibinfo {pages} {512}\BibitemShut {NoStop}%
\bibitem [{\citenamefont {Hanwell}\ \emph {et~al.}(2012)\citenamefont
  {Hanwell}, \citenamefont {Curtis}, \citenamefont {Lonie}, \citenamefont
  {Vandermeersch}, \citenamefont {Zurek},\ and\ \citenamefont
  {Hutchison}}]{Hanwell2012}%
  \BibitemOpen
  \bibfield  {author} {\bibinfo {author} {\bibfnamefont {M.~D.}\ \bibnamefont
  {Hanwell}}, \bibinfo {author} {\bibfnamefont {D.~E.}\ \bibnamefont {Curtis}},
  \bibinfo {author} {\bibfnamefont {D.~C.}\ \bibnamefont {Lonie}}, \bibinfo
  {author} {\bibfnamefont {T.}~\bibnamefont {Vandermeersch}}, \bibinfo {author}
  {\bibfnamefont {E.}~\bibnamefont {Zurek}}, \ and\ \bibinfo {author}
  {\bibfnamefont {G.~R.}\ \bibnamefont {Hutchison}},\ }\href {\doibase
  10.1186/1758-2946-4-17} {\bibfield  {journal} {\bibinfo  {journal} {J.
  Cheminf.}\ }\textbf {\bibinfo {volume} {4}},\ \bibinfo {pages} {17} (\bibinfo
  {year} {2012})}\BibitemShut {NoStop}%
\bibitem [{\citenamefont {Flyvbjerg}\ and\ \citenamefont
  {Petersen}(1989)}]{Flyvbjerg1989}%
  \BibitemOpen
  \bibfield  {author} {\bibinfo {author} {\bibfnamefont {H.}~\bibnamefont
  {Flyvbjerg}}\ and\ \bibinfo {author} {\bibfnamefont {H.~G.}\ \bibnamefont
  {Petersen}},\ }\href {\doibase 10.1063/1.457480} {\bibfield  {journal}
  {\bibinfo  {journal} {J. Chem. Phys.}\ }\textbf {\bibinfo {volume} {91}},\
  \bibinfo {pages} {461} (\bibinfo {year} {1989})}\BibitemShut {NoStop}%
\bibitem [{\citenamefont {Frenkel}\ and\ \citenamefont
  {Smit}(2001)}]{Frenkel2001}%
  \BibitemOpen
  \bibfield  {author} {\bibinfo {author} {\bibfnamefont {D.}~\bibnamefont
  {Frenkel}}\ and\ \bibinfo {author} {\bibfnamefont {B.}~\bibnamefont {Smit}},\
  }\href@noop {} {\emph {\bibinfo {title} {{Understanding Molecular
  Simulations: from Algorithms to Application}}}},\ \bibinfo {edition} {2nd}\
  ed.\ (\bibinfo  {publisher} {Elsevier},\ \bibinfo {year} {2001})\BibitemShut
  {NoStop}%
\bibitem [{\citenamefont {Kumar}\ \emph {et~al.}(2007)\citenamefont {Kumar},
  \citenamefont {Schmidt},\ and\ \citenamefont {Skinner}}]{Kumar2007}%
  \BibitemOpen
  \bibfield  {author} {\bibinfo {author} {\bibfnamefont {R.}~\bibnamefont
  {Kumar}}, \bibinfo {author} {\bibfnamefont {J.~R.}\ \bibnamefont {Schmidt}},
  \ and\ \bibinfo {author} {\bibfnamefont {J.~L.}\ \bibnamefont {Skinner}},\
  }\href {\doibase 10.1063/1.2742385} {\bibfield  {journal} {\bibinfo
  {journal} {J. Chem. Phys.}\ }\textbf {\bibinfo {volume} {126}},\ \bibinfo
  {pages} {204107} (\bibinfo {year} {2007})}\BibitemShut {NoStop}%
\bibitem [{\citenamefont {Kathmann}\ \emph
  {et~al.}(2007{\natexlab{a}})\citenamefont {Kathmann}, \citenamefont
  {Schenter},\ and\ \citenamefont {Garrett}}]{Kathmann2007}%
  \BibitemOpen
  \bibfield  {author} {\bibinfo {author} {\bibfnamefont {S.}~\bibnamefont
  {Kathmann}}, \bibinfo {author} {\bibfnamefont {G.}~\bibnamefont {Schenter}},
  \ and\ \bibinfo {author} {\bibfnamefont {B.}~\bibnamefont {Garrett}},\ }\href
  {http://prl.aps.org/abstract/PRL/v98/i10/e109603} {\bibfield  {journal}
  {\bibinfo  {journal} {Phys. Rev. Lett.}\ }\textbf {\bibinfo {volume} {98}},\
  \bibinfo {pages} {109603} (\bibinfo {year} {2007}{\natexlab{a}})}\BibitemShut
  {NoStop}%
\bibitem [{\citenamefont {Kathmann}\ \emph
  {et~al.}(2007{\natexlab{b}})\citenamefont {Kathmann}, \citenamefont
  {Schenter},\ and\ \citenamefont {Garrett}}]{Kathmann2007b}%
  \BibitemOpen
  \bibfield  {author} {\bibinfo {author} {\bibfnamefont {S.}~\bibnamefont
  {Kathmann}}, \bibinfo {author} {\bibfnamefont {G.}~\bibnamefont {Schenter}},
  \ and\ \bibinfo {author} {\bibfnamefont {B.}~\bibnamefont {Garrett}},\ }\href
  {\doibase 10.1021/jp067468u} {\bibfield  {journal} {\bibinfo  {journal} {J.
  Phys. Chem. C}\ }\textbf {\bibinfo {volume} {111}},\ \bibinfo {pages} {4977}
  (\bibinfo {year} {2007}{\natexlab{b}})}\BibitemShut {NoStop}%
\bibitem [{\citenamefont {Kurt\'{e}n}\ \emph {et~al.}(2008)\citenamefont
  {Kurt\'{e}n}, \citenamefont {Loukonen}, \citenamefont {Vehkam\"{a}ki},\ and\
  \citenamefont {Kulmala}}]{Kurten2008}%
  \BibitemOpen
  \bibfield  {author} {\bibinfo {author} {\bibfnamefont {T.}~\bibnamefont
  {Kurt\'{e}n}}, \bibinfo {author} {\bibfnamefont {V.}~\bibnamefont
  {Loukonen}}, \bibinfo {author} {\bibfnamefont {H.}~\bibnamefont
  {Vehkam\"{a}ki}}, \ and\ \bibinfo {author} {\bibfnamefont {M.}~\bibnamefont
  {Kulmala}},\ }\href {\doibase 10.5194/acpd-8-7455-2008} {\bibfield  {journal}
  {\bibinfo  {journal} {Atmospheric Chemistry and Physics Discussions}\
  }\textbf {\bibinfo {volume} {8}},\ \bibinfo {pages} {7455} (\bibinfo {year}
  {2008})}\BibitemShut {NoStop}%
\bibitem [{\citenamefont {Kirkby}\ \emph {et~al.}(2011)\citenamefont {Kirkby},
  \citenamefont {Curtius}, \citenamefont {Almeida}, \citenamefont {Dunne},
  \citenamefont {Duplissy}, \citenamefont {Ehrhart}, \citenamefont {Franchin},
  \citenamefont {Gagn\'{e}}, \citenamefont {Ickes}, \citenamefont {K\"{u}rten},
  \citenamefont {Kupc}, \citenamefont {Metzger}, \citenamefont {Riccobono},
  \citenamefont {Rondo}, \citenamefont {Schobesberger}, \citenamefont
  {Tsagkogeorgas}, \citenamefont {Wimmer}, \citenamefont {Amorim},
  \citenamefont {Bianchi}, \citenamefont {Breitenlechner}, \citenamefont
  {David}, \citenamefont {Dommen}, \citenamefont {Downard}, \citenamefont
  {Ehn}, \citenamefont {Flagan}, \citenamefont {Haider}, \citenamefont
  {Hansel}, \citenamefont {Hauser}, \citenamefont {Jud}, \citenamefont
  {Junninen}, \citenamefont {Kreissl}, \citenamefont {Kvashin}, \citenamefont
  {Laaksonen}, \citenamefont {Lehtipalo}, \citenamefont {Lima}, \citenamefont
  {Lovejoy}, \citenamefont {Makhmutov}, \citenamefont {Mathot}, \citenamefont
  {Mikkil\"{a}}, \citenamefont {Minginette}, \citenamefont {Mogo},
  \citenamefont {Nieminen}, \citenamefont {Onnela}, \citenamefont {Pereira},
  \citenamefont {Pet\"{a}j\"{a}}, \citenamefont {Schnitzhofer}, \citenamefont
  {Seinfeld}, \citenamefont {Sipil\"{a}}, \citenamefont {Stozhkov},
  \citenamefont {Stratmann}, \citenamefont {Tom\'{e}}, \citenamefont
  {Vanhanen}, \citenamefont {Viisanen}, \citenamefont {Vrtala}, \citenamefont
  {Wagner}, \citenamefont {Walther}, \citenamefont {Weingartner}, \citenamefont
  {Wex}, \citenamefont {Winkler}, \citenamefont {Carslaw}, \citenamefont
  {Worsnop}, \citenamefont {Baltensperger},\ and\ \citenamefont
  {Kulmala}}]{Kirkby2011}%
  \BibitemOpen
  \bibfield  {author} {\bibinfo {author} {\bibfnamefont {J.}~\bibnamefont
  {Kirkby}}, \bibinfo {author} {\bibfnamefont {J.}~\bibnamefont {Curtius}},
  \bibinfo {author} {\bibfnamefont {J.}~\bibnamefont {Almeida}}, \bibinfo
  {author} {\bibfnamefont {E.}~\bibnamefont {Dunne}}, \bibinfo {author}
  {\bibfnamefont {J.}~\bibnamefont {Duplissy}}, \bibinfo {author}
  {\bibfnamefont {S.}~\bibnamefont {Ehrhart}}, \bibinfo {author} {\bibfnamefont
  {A.}~\bibnamefont {Franchin}}, \bibinfo {author} {\bibfnamefont
  {S.}~\bibnamefont {Gagn\'{e}}}, \bibinfo {author} {\bibfnamefont
  {L.}~\bibnamefont {Ickes}}, \bibinfo {author} {\bibfnamefont
  {A.}~\bibnamefont {K\"{u}rten}}, \bibinfo {author} {\bibfnamefont
  {A.}~\bibnamefont {Kupc}}, \bibinfo {author} {\bibfnamefont {A.}~\bibnamefont
  {Metzger}}, \bibinfo {author} {\bibfnamefont {F.}~\bibnamefont {Riccobono}},
  \bibinfo {author} {\bibfnamefont {L.}~\bibnamefont {Rondo}}, \bibinfo
  {author} {\bibfnamefont {S.}~\bibnamefont {Schobesberger}}, \bibinfo {author}
  {\bibfnamefont {G.}~\bibnamefont {Tsagkogeorgas}}, \bibinfo {author}
  {\bibfnamefont {D.}~\bibnamefont {Wimmer}}, \bibinfo {author} {\bibfnamefont
  {A.}~\bibnamefont {Amorim}}, \bibinfo {author} {\bibfnamefont
  {F.}~\bibnamefont {Bianchi}}, \bibinfo {author} {\bibfnamefont
  {M.}~\bibnamefont {Breitenlechner}}, \bibinfo {author} {\bibfnamefont
  {A.}~\bibnamefont {David}}, \bibinfo {author} {\bibfnamefont
  {J.}~\bibnamefont {Dommen}}, \bibinfo {author} {\bibfnamefont
  {A.}~\bibnamefont {Downard}}, \bibinfo {author} {\bibfnamefont
  {M.}~\bibnamefont {Ehn}}, \bibinfo {author} {\bibfnamefont {R.~C.}\
  \bibnamefont {Flagan}}, \bibinfo {author} {\bibfnamefont {S.}~\bibnamefont
  {Haider}}, \bibinfo {author} {\bibfnamefont {A.}~\bibnamefont {Hansel}},
  \bibinfo {author} {\bibfnamefont {D.}~\bibnamefont {Hauser}}, \bibinfo
  {author} {\bibfnamefont {W.}~\bibnamefont {Jud}}, \bibinfo {author}
  {\bibfnamefont {H.}~\bibnamefont {Junninen}}, \bibinfo {author}
  {\bibfnamefont {F.}~\bibnamefont {Kreissl}}, \bibinfo {author} {\bibfnamefont
  {A.}~\bibnamefont {Kvashin}}, \bibinfo {author} {\bibfnamefont
  {A.}~\bibnamefont {Laaksonen}}, \bibinfo {author} {\bibfnamefont
  {K.}~\bibnamefont {Lehtipalo}}, \bibinfo {author} {\bibfnamefont
  {J.}~\bibnamefont {Lima}}, \bibinfo {author} {\bibfnamefont {E.~R.}\
  \bibnamefont {Lovejoy}}, \bibinfo {author} {\bibfnamefont {V.}~\bibnamefont
  {Makhmutov}}, \bibinfo {author} {\bibfnamefont {S.}~\bibnamefont {Mathot}},
  \bibinfo {author} {\bibfnamefont {J.}~\bibnamefont {Mikkil\"{a}}}, \bibinfo
  {author} {\bibfnamefont {P.}~\bibnamefont {Minginette}}, \bibinfo {author}
  {\bibfnamefont {S.}~\bibnamefont {Mogo}}, \bibinfo {author} {\bibfnamefont
  {T.}~\bibnamefont {Nieminen}}, \bibinfo {author} {\bibfnamefont
  {A.}~\bibnamefont {Onnela}}, \bibinfo {author} {\bibfnamefont
  {P.}~\bibnamefont {Pereira}}, \bibinfo {author} {\bibfnamefont
  {T.}~\bibnamefont {Pet\"{a}j\"{a}}}, \bibinfo {author} {\bibfnamefont
  {R.}~\bibnamefont {Schnitzhofer}}, \bibinfo {author} {\bibfnamefont {J.~H.}\
  \bibnamefont {Seinfeld}}, \bibinfo {author} {\bibfnamefont {M.}~\bibnamefont
  {Sipil\"{a}}}, \bibinfo {author} {\bibfnamefont {Y.}~\bibnamefont
  {Stozhkov}}, \bibinfo {author} {\bibfnamefont {F.}~\bibnamefont {Stratmann}},
  \bibinfo {author} {\bibfnamefont {A.}~\bibnamefont {Tom\'{e}}}, \bibinfo
  {author} {\bibfnamefont {J.}~\bibnamefont {Vanhanen}}, \bibinfo {author}
  {\bibfnamefont {Y.}~\bibnamefont {Viisanen}}, \bibinfo {author}
  {\bibfnamefont {A.}~\bibnamefont {Vrtala}}, \bibinfo {author} {\bibfnamefont
  {P.~E.}\ \bibnamefont {Wagner}}, \bibinfo {author} {\bibfnamefont
  {H.}~\bibnamefont {Walther}}, \bibinfo {author} {\bibfnamefont
  {E.}~\bibnamefont {Weingartner}}, \bibinfo {author} {\bibfnamefont
  {H.}~\bibnamefont {Wex}}, \bibinfo {author} {\bibfnamefont {P.~M.}\
  \bibnamefont {Winkler}}, \bibinfo {author} {\bibfnamefont {K.~S.}\
  \bibnamefont {Carslaw}}, \bibinfo {author} {\bibfnamefont {D.~R.}\
  \bibnamefont {Worsnop}}, \bibinfo {author} {\bibfnamefont {U.}~\bibnamefont
  {Baltensperger}}, \ and\ \bibinfo {author} {\bibfnamefont {M.}~\bibnamefont
  {Kulmala}},\ }\href {\doibase 10.1038/nature10343} {\bibfield  {journal}
  {\bibinfo  {journal} {Nature}\ }\textbf {\bibinfo {volume} {476}},\ \bibinfo
  {pages} {429} (\bibinfo {year} {2011})}\BibitemShut {NoStop}%
\end{thebibliography}

\end{document}